\documentclass[journal]{IEEEtran}
\usepackage{amssymb}
\usepackage{cite,graphicx,amsmath,amssymb}
\usepackage{ifpdf, ifxetex}
\ifxetex
\usepackage{graphicx}
\else
\ifpdf
\usepackage{graphicx}
\usepackage{epstopdf}
\else
\usepackage[dvipdfm]{graphicx}
\usepackage{bmpsize}
\fi
\fi

\usepackage{amsmath}
\usepackage{mathrsfs}
\usepackage{amsthm}
\usepackage{mathrsfs} 
\usepackage{algorithm}
\usepackage{algorithmic}
\usepackage{bm}
\usepackage{graphicx} 
\usepackage{subfigure}
\usepackage{pifont}
\usepackage{latexsym}
\usepackage{amsfonts}
\usepackage{color}
\usepackage{xcolor}
\usepackage{amssymb}

\usepackage{array}
\usepackage{stfloats} 
\usepackage{url}

\usepackage{paralist}
\usepackage{mathtools}
\usepackage{xcolor}
\usepackage{xpatch}
\usepackage[english]{babel}

\usepackage{cite}
\usepackage{algorithm}
\usepackage{algorithmic}
\usepackage{multirow}
\usepackage{caption}
\begin{document}

\title{Rate-Splitting Multiple Access Enabled Probabilistic Semantic Communication in UAV Networks}
\author{Sicheng~Wang, ~\IEEEmembership{Graduate Student Member,~IEEE,}
        Tiankui~Zhang,~\IEEEmembership{Senior Member,~IEEE,}
        Xu~Gan,~\IEEEmembership{Member,~IEEE,}
        and Wenjun~Xu,~\IEEEmembership{Senior Member,~IEEE}
\thanks{Sicheng Wang, Tiankui Zhang and Wenjun Xu are with the School of Information and Communication Engineering, Beijing University of Posts and Telecommunications, Beijing, 100876, China (e-mail: wangsicheng@bupt.edu.cn; zhangtiankui@bupt.edu.cn; wjxu@bupt.edu.cn).}
\thanks{Xu~Gan is with the Department of Electrical and Electronic Engineering, The University of Hong Kong, Hong Kong (e-mail: eee.ganxu@hku.hk)}
}

\maketitle

\begin{abstract}
This article proposes an uncrewed aerial vehicle (UAV) downlink semantic communication framework, where probabilistic knowledge graphs (PKGs) are employed to model user equipment (UE) semantics and decompose semantic information into shared and private components.
Leveraging the capability of rate-splitting multiple access (RSMA) in addressing such semantic structures, a PKG-assisted RSMA transmission scheme is developed to efficiently deliver multi-user semantic information under severe energy constraints and fast-varying UAV channels.
To characterize the strongly coupled energy costs of communication, computation, and flight, a weighted energy minimization problem is formulated to jointly optimize the UAV trajectory, power allocation, beamforming design, and semantic compression ratio.
The resulting non-convex problem is efficiently solved using an iterative semantic-aware weighted energy optimization (SWEO) algorithm that integrates Lagrangian dual decomposition and successive convex approximation. Furthermore, a semantic accuracy metric is proposed to quantify the reliability of reconstruction by assigning importance-based weights to informative KG triples.
Extensive simulation results verify that the proposed framework achieves superior energy efficiency, enhanced semantic preservation, and consistently better performance than conventional RSMA, non-orthogonal multiple access (NOMA), and space division multiple access (SDMA) schemes in benchmarks across various network parameters.
\end{abstract}


\section{Introduction}
The sixth-generation (6G) network is expected to be a paradigm shift in mobile wireless communications, enabling emerging applications with stringent requirements on ultra-low latency, high data rates, and massive connectivity~\cite{6G1,6G2}. 
Uncrewed aerial vehicles (UAVs) offer distinct advantages in terms of flexible deployment and robust line-of-sight (LoS) links, making them a promising solution to support such demanding applications in scenarios such as disaster relief, emergency communications, and remote area coverage~\cite{UAV1,UAV2}.
However, UAV networks are fundamentally constrained by limited onboard energy resources, where communication, computation, and propulsion jointly incur significant energy consumption burden~\cite{Challenge2}.

The stringent energy constraints of UAV networks highlight the need for more efficient transmission schemes~\cite{FAS}. Recent advances in wireless edge artificial intelligence suggest that future communication systems should move beyond conventional bit-level transmission and focus on delivering task-relevant information, thereby enabling efficient intelligent services at the network edge~\cite{AI}. Fortunately, semantic communication utilizes task-oriented information and eliminates redundancy to address the UAV energy constraints.
Unlike traditional bit-oriented communication, semantic communication extracts relevant information based on the communication task and filters out redundant data. 
In practice, semantic communication has already been applied to various transmission tasks and shown significant performance gains, including text~\cite{text}, speech~\cite{speech}, image~\cite{image}, and video~\cite{video}. 
Despite its potential in reducing redundancy, semantic communication fundamentally relies on the shared understanding of contextual knowledge between the transmitter and the receiver~\cite{Challenge3}.
However, in dynamic UAV communications, maintaining knowledge consistency is difficult. 
To overcome this challenge, knowledge graphs (KGs) have attracted increasing attention in recent years~\cite{RSMA5, KGa, KGb}. Specifically, KGs serve as structured representations of abstract facts \cite{KG1}. 
By structuring information into triples: (\textit{entity, relation, entity}), KGs provide a machine-understandable framework for semantic representation and reasoning, thereby offering a promising approach to building more intelligent and accurate semantic communication systems~\cite{KG2}.
Recent studies have explored the integration of KGs into semantic communication frameworks~\cite{KG_1, KG_2, KG_3}.
Jiang et al.~\cite{KG_1} converted the transmitted sentences into triples of a KG and adaptively transmitted content according to semantic importance and channel quality, thereby improving the communication reliability in low signal-to-noise ratio regimes.
Zhang et al.~\cite{KG_2} proposed a KG-based semantic communication framework for cooperative image transmission and achieved semantic-oriented resource allocation, significantly reducing transmission delay.
Zhou et al.~\cite{KG_3} developed KG-based cognitive semantic communication frameworks with interpretable semantic alignment and correction algorithms, enhancing both compression efficiency and communication reliability.
While the aforementioned studies~\cite{KG_1, KG_2, KG_3} have advanced KG-based semantic communication, the multi-user characteristics of KGs remain insufficiently investigated, and inter-user semantic correlations are not fully exploited.
Actually, in practical semantic communication scenarios, different users may share overlapping semantic entities within their transmitted knowledge graphs.
Motivated by this observation, we decompose each user’s information into shared and private components, and employ rate-splitting multiple access (RSMA) to transmit them efficiently in this work. 

RSMA has recently emerged as a promising non-orthogonal multiple access (NOMA) technique for 6G wireless communications~\cite{RSMA1, RSMA2, RSMA3}.
By enabling partial decoding of interference through rate splitting, RSMA provides a unified framework that generalizes and outperforms conventional NOMA and space division multiple access (SDMA) in terms of spectral and energy efficiency~\cite{RSMA4,RSMA5}. 
Owing to these advantages, RSMA has been widely adopted in various emerging communication paradigms, such as reconfigurable intelligent surface (RIS)-assisted systems~\cite{RIS}, cognitive satellite-terrestrial networks~\cite{STN}, and UAV communications~\cite{RSMAUAV}.
Meanwhile, with the recent rise of semantic communication as an efficient paradigm for meaning-level information transmission, researchers have begun exploring the integration of RSMA into semantic communication frameworks.
In multi-user semantic communication scenarios, users may naturally share overlapping semantic content or interests.
This observation has motivated researchers to incorporate RSMA into semantic communication systems, leveraging its capability to flexibly split shared and private information for efficient multi-user semantic transmission~\cite{RSMASem1, RSMASem2, RSMASem3, RSMASem4}.
Yang et al.~\cite{RSMASem1} optimized joint computation and communication resources in an RSMA-enabled semantic communication framework to minimize system energy consumption.
Zeng et al.~\cite{RSMASem2} applied RSMA to semantic communication in wireless control systems to enhance semantic transmission efficiency under strict latency constraints.
Liu et al.~\cite{RSMASem3} further investigated RSMA for uplink coexistence of semantic and bit communications in 6G networks, demonstrating its superiority in mixed transmission scenarios.
Cheng et al.~\cite{RSMASem4} developed an RSMA-enabled task-oriented semantic image transmission framework that optimizes quality of experience (QoE) through joint power and message selection strategies, outperforming SDMA benchmarks.

Nevertheless, existing RSMA-based semantic communication works are mainly confined to static or terrestrial scenarios, and do not consider UAV-enabled systems where channel conditions, energy consumption, and semantic transmission efficiency are jointly affected by mobility.
Moreover, conventional schemes often treat all extracted semantic features equally, lacking an interpretable metric to quantitatively evaluate the preservation of critical semantic knowledge.
In contrast, UAV semantic communication introduces trajectory-dependent probabilistic line-of-sight (LoS)/non-line-of-sight (NLoS) channels, where UAV mobility directly affects both communication quality and propulsion energy consumption. Meanwhile, probabilistic knowledge graph (PKG)-based semantic compression introduces additional tradeoffs among computation load, communication cost, and semantic reconstruction accuracy. Under these coupled effects, the joint optimization of semantic compression, RSMA resource allocation, and UAV trajectory design becomes a challenging and largely unexplored problem.

To address the above challenges, we develop a UAV semantic communication framework based on PKGs, which enables downlink transmission from an airborne mobile base station (BS) to terrestrial user equipments (UEs). 
In particular, the semantic information of different UEs consists of both shared and private components, which makes RSMA a natural choice.
Under this framework, the energy-efficient transmission of multi-user probabilistic semantic communication (PSC) in UAV networks is investigated.
The contributions of this work are summarized as follows:

\begin{itemize}  
    \item 
    We propose a UAV probabilistic semantic communication framework for downlink transmission, where shared and private semantic information are flexibly delivered using RSMA. A PKG-assisted semantic compression scheme is developed, in which the semantic compression ratio is introduced as a tunable variable to balance computation overhead and communication efficiency. Moreover, a semantic accuracy metric is introduced to quantitatively evaluate the preservation of essential semantic information by assigning higher weights to more informative knowledge graph triples.
    \item 
    We formulate a weighted energy consumption problem which jointly accounts for communication, computation, and flight energy in UAV networks. The objective is to minimize the overall weighted energy consumption through coordinated optimization of the UAV trajectory, power allocation, beamforming, and semantic compression ratio. To efficiently address the resulting complex non-convex problem, we propose a semantic-aware weighted energy optimization (SWEO) algorithm. The algorithm addresses the involved subproblems in an alternating manner, leveraging Lagrangian dual decomposition and successive convex approximation (SCA) techniques.
    \item 
    We demonstrate the effectiveness of the proposed framework and optimization algorithm through extensive numerical simulations under dynamic UAV communication scenarios.
    Simulation results demonstrate that: 
    1) the proposed RSMA-enabled PSC transmission scheme consistently outperforms conventional RSMA, SDMA and NOMA transmission scheme under various bandwidth, transmission data sizes and SNR. 
    2) The proposed weighted semantic accuracy metric better captures the loss of critical semantics compared with the conventional unweighted metric, thus leading to improved task-level performance. 
    3) The proposed algorithm outperforms baseline schemes by achieving an optimal trade-off among communication, computation, and flight energy consumption.
\end{itemize}

The remainder of this paper is organized as follows. 
The system model of the UAV semantic communication system model is illustrated in Section II. 
Moreover, Section III formulates the weight energy consumption minimization problem and develops iterative alternating optimization  algorithm.
Simulation results are demonstrated in Section IV. 
Conclusions are drawn in Section V.
\begin{table}[!t]
\small
\centering
\caption{Summary of Notations}
\label{tab2-1} 
\begin{tabular}{|>{\arraybackslash}p{1.6cm}|>{\arraybackslash}p{5.7cm}|}
            \hline
                \textbf{Notation}         & \textbf{Definition} \\
            \hline
                $K$                & Total number of UEs \\
            \hline
                $T_v$              & Total UAV flight time \\
            \hline
                $T$                & Total number of time slots \\
            \hline
                $\delta$           & Length of each time slot \\
            \hline
                $V_{\max}$         & Maximum flight speed of the UAV\\
            \hline
                $H$                & Altitude of the UAV \\
            \hline
                $\mathbf{q}$             & Position of the UAV \\
            \hline
                $\mathbf{L}_k$      & Position of the UE $k$\\
            \hline
                $d_k$   & Distances between the UAV and UE $k$ \\
            \hline
                $\mu_k$     & Path loss between the UAV and UE $k$\\
            \hline
                $P^\text{LoS}_{k},P^\text{NLoS}_{k}$ & Probabilities of LoS and NLoS link states \\
            \hline
                $\varphi_{k}$ & Elevation angle between the UAV and UE $k$\\
            \hline
                $h_k$   & Channel gains between the UAV and UE $k$\\
            \hline
                $n_k$              & Additive white Gaussian noise at UE $k$ \\
            \hline
                $u_0, u_k$         & Number of shared and private triples \\
            \hline
                \multirow{2}{*}{$\Omega_0, \Omega_k$} & Compression ratios for the shared and private components\\
            \hline
                $\tau_0, \tau_k$ & Delay of shared and private information\\
            \hline
                $L_e$ & Bit-length required to represent an entity \\
            \hline
                \multirow{2}{*}{$f_0, f_k$} & Computation capabilities allocated for semantic compression \\
            \hline
                $\widetilde{\Gamma}_k^{\rm W}$ & PKG-weighted semantic accuracy metric\\
            \hline
                \multirow{2}{*}{$\mathbf{w}_0, \mathbf{w}_k$} & Beamforming vectors for the shared and private information \\
            \hline
                \multirow{2}{*}{$p_0, p_k$}   & Transmit powers for the shared and private streams\\
            \hline
                \multirow{2}{*}{$R^s_k, R^p_k$}  &  Achievable shared and private transmission rates for UE $k$\\
            \hline
                \multirow{2}{*}{$E_1, E_2, E_3$} & Computation, communication, and UAV flight energy consumption\\
            \hline
                $\omega$ & Weight factor for energy consumption\\
            \hline
\end{tabular}   
\end{table}
\section{System Model}
\begin{figure}[!t]
    \centering
    \includegraphics[width=\columnwidth]{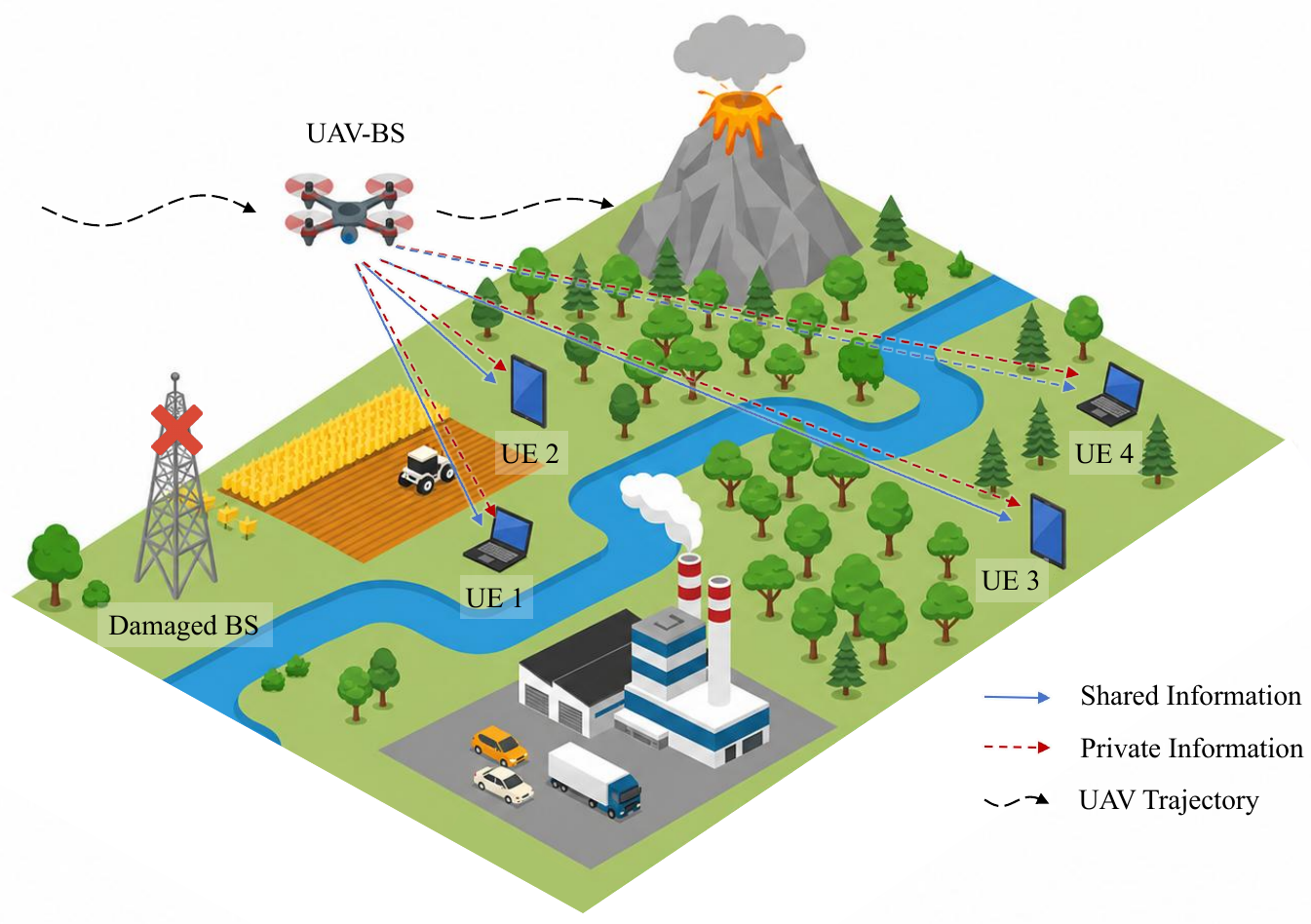}
    \caption{Illustration of RSMA-enabled UAV probabilistic semantic communications framework.}
    \label{Communication scenario}
\end{figure}
As shown in~Fig.~\ref{Communication scenario}, the UAV-BS equipped with $M$ antennas provides downlink communication services to $K$ single-antenna UEs. The set of users is denoted by $\mathcal{K} = \{1, 2, \ldots, K\}$.
The UAV-BS aims to transmit the task-related data set $\mathcal{D}_k$ to each UE $k$.
The complete data set can be represented as $\mathcal{D} = \{\mathcal{D}_1, \mathcal{D}_2, \dots, \mathcal{D}_K\}$.
The original data $\mathcal{D}_k$ is processed to obtain compact semantic representations $\mathcal{S}_k$, which are then divided into shared information $\mathcal{U}_0$ and private information $\{\mathcal{U}_1, \mathcal{U}_2, \dots, \mathcal{U}_K\}$.
Both shared and private information are further compressed into the semantic features based on the PKG and then encoded for downlink RSMA transmission.
Table \ref{tab2-1} provides a summary of the symbols used in this paper.

\subsection{Semantic Communication Model}
\begin{figure}[!t]
    \centering
    \includegraphics[width=\columnwidth]{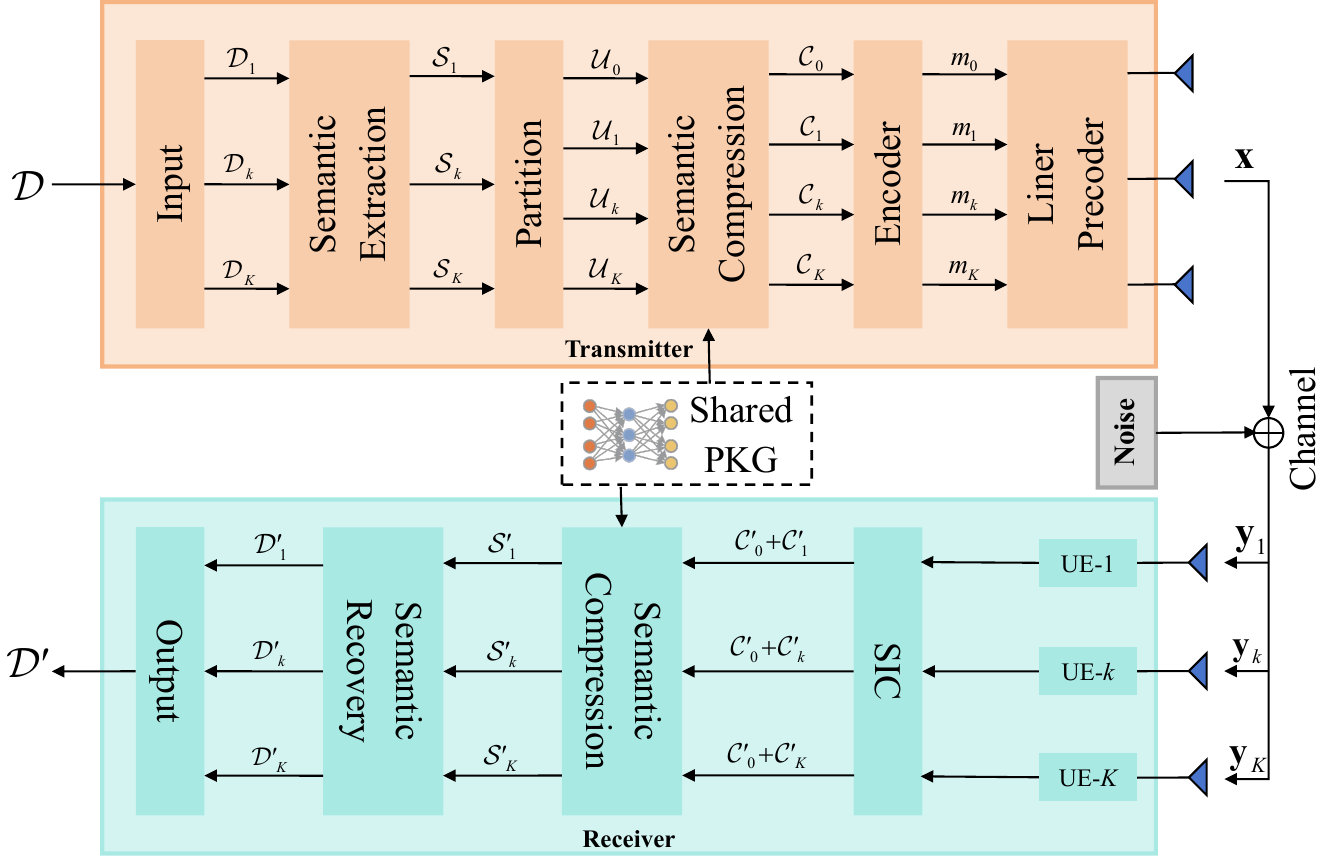}
    \caption{Block diagram of RSMA-enabled probabilistic semantic communications.}
    \label{System_Model}
\end{figure}
This paper introduces a probabilistic semantic communication system into UAV networks~\cite{RSMA5}.
As illustrated in Fig.~\ref{System_Model}, the UAV performs a semantic extraction process on the data set $\mathcal{D}$ based on the locally obtained knowledge base.
In this work, a consistent PKG shared by the UAV and UEs is assumed to be established beforehand to focus on the semantic transmission process. In practice, the PKG can be constructed offline from historical data and periodically updated to accommodate environmental changes. To reduce synchronization overhead, only lightweight knowledge updates need to be disseminated among the UAV and UEs.
Specifically, semantic triples are extracted to form the semantic representations of all UEs, denoted by $\mathcal{S}=\{\mathcal{S}_1,\mathcal{S}_2,\ldots,\mathcal{S}_K\}$.
The semantic extraction process can be expressed as
\begin{equation}
    \mathcal{S}=T_{\phi}(\mathcal{D}),
    \label{eq:semantic_extraction}
\end{equation}
where $T_{\phi}(\cdot)$ denotes a pre-trained semantic extractor parameterized by $\phi$, which converts the raw dataset into semantic triples for subsequent PKG construction and semantic partitioning.

After semantic extraction, the semantic information of all UEs is analyzed to identify both shared and UE-specific triples.
Specifically, after KG alignment, a triple is assigned to the shared semantic set if it is semantically matched across all UE semantic sets
\begin{equation}
    \mathcal{U}_0 = \bigcap_{k=1}^{K} \mathcal{S}_k .
    \label{eq:shared_set}
\end{equation}
The remaining triples are treated as UE-specific semantic information and transmitted through private streams.
Therefore, the private semantic set of UE $k$ is defined as
\begin{equation}
    \mathcal{U}_k = \mathcal{S}_k \setminus \mathcal{U}_0,
    \quad \forall k\in\mathcal{K}.
    \label{eq:private_set}
\end{equation}
Accordingly, the semantic partitioning process can be expressed as
\begin{equation}
    \mathcal{U}=F(\mathcal{S}),
    \label{eq:semantic_partition}
\end{equation}
where $F(\cdot)$ denotes the semantic partitioning operator that divides the extracted semantic sets into the shared semantic component $\mathcal{U}_0$ and private semantic components $\{\mathcal{U}_k\}_{k=1}^{K}$ according to the above criterion.

To configure the RSMA transmission structure, the strict-intersection criterion defined in \eqref{eq:shared_set}-\eqref{eq:private_set} is adopted to separate shared and UE-specific semantic information. Since the RSMA common stream is decoded by all UEs, this criterion ensures that the common stream carries only semantic information required by every UE, thereby avoiding unnecessary decoding overhead. Although semantic overlap in practical systems may be partial rather than universal, accommodating such partial overlap would substantially increase the complexity of the RSMA transmission framework and the associated joint optimization problem. Therefore, the adopted criterion provides a tractable semantic partitioning strategy that remains fully consistent with the single-common-stream RSMA architecture considered in this paper.

The UAV maintains a local PKG constructed from the knowledge base.
Each triple in the PKG is associated with a probabilistic weight $\kappa$ representing its statistical importance. 
Based on the PKG, semantic compression is performed to obtain the compact representation,
\begin{equation}
    \mathcal{C}=G_{\xi}(\mathcal{U},\kappa),
    \label{eq:semantic_compression}
\end{equation}
where $G_{\xi}(\cdot)$ denotes the semantic compression operator parameterized by the compression strategy $\xi$.
The resulting compressed semantic representation is denoted by
$\mathcal{C}=\{\mathcal{C}_0,\mathcal{C}_1,\ldots,\mathcal{C}_K\}$,
where $\mathcal{C}_0$ corresponds to the compressed shared semantic component and $\mathcal{C}_k$ corresponds to the compressed private semantic component of UE $k$.
The compressed semantic set $\mathcal{C}$ is then encoded before transmission as
\begin{equation}
    m = E_{\eta}(\mathcal{C}),
    \label{eq:semantic_encoding}
\end{equation}
where $E_{\eta}(\cdot)$ denotes the semantic encoder parameterized by $\eta$.
The resulting encoded symbol stream is given by
$m=\{m_0,m_1,\ldots,m_K\}$,
where $m_0$ corresponds to the encoded shared information generated from $\mathcal{C}_0$, and $m_k$ represents the encoded private information generated from $\mathcal{C}_k$ for UE $k$.

At the receiver, the UEs sequentially decode the shared and private information.
Specifically, each UE first decodes the common stream and then decodes its private stream using SIC.
The recovered compressed semantic representation of UE $k$ can be expressed as $\mathcal{C}'_k \cup \mathcal{C}'_0$, where $\mathcal{C}'_0$ and $\mathcal{C}'_k$ denote the decoded shared semantic component and private semantic component for UE $k$, respectively.
Therefore, the recovered compressed semantic representations of all UEs can be expressed as $\mathcal{C}' = \{\mathcal{C}'_1 \cup \mathcal{C}'_0, \dots, \mathcal{C}'_k \cup \mathcal{C}'_0, \dots, \mathcal{C}'_K\cup \mathcal{C}'_0\}$.
Subsequently, based on the shared PKG and the local decoding process, each UE performs semantic reconstruction,
\begin{equation}
    \mathcal{S}' = R_{\varphi}(\mathcal{C}', \kappa),
    \label{eq:semantic_reconstruction}
\end{equation}
where $\mathcal{S}'$ denotes the recovered semantic representations,  $R_{\varphi}(\cdot)$ denotes the semantic reconstruction operator parameterized by $\varphi$, and $\kappa$ represents the probabilistic correlation parameters from the shared PKG utilized during decoding.

Finally, the reconstructed semantic representations are transformed back into the task domain through semantic recovery, yielding the reconstructed data
\begin{equation}
    \mathcal{D}' = D_{\theta}(\mathcal{S}'),
    \label{eq:semantic_recovery}
\end{equation}
where $D_{\theta}(\cdot)$ denotes the semantic decoder parameterized by $\theta$.
The recovered data $\mathcal{D}' = \{\mathcal{D}'_1, \mathcal{D}'_2, \dots, \mathcal{D}'_K\}$ correspond to the estimated task-related information for each UE.

The semantic processing functions $T_{\phi}(\cdot)$, $G_{\xi}(\cdot)$, $E_{\eta}(\cdot)$, $R_{\varphi}(\cdot)$, and $D_{\theta}(\cdot)$ represent task-oriented semantic modules that are trained or constructed offline before communication.
The proposed optimization framework does not optimize the neural parameters $\phi$, $\xi$, $\eta$, $\varphi$, and $\theta$, but focuses on the online optimization of semantic compression ratio, RSMA transmission resources, and UAV trajectory based on the extracted and compressed semantic representations.

To characterize the statistical regularity of semantic triples, the PKG-assisted probabilistic semantic compression principle in~\cite{RSMA5} is adopted as the semantic-layer modeling basis. The resulting model provides a tractable interface between semantic compression, computation cost, and communication load in the considered RSMA-enabled UAV network.

Assume that the UAV-BS has constructed a local KG from a task-related sample dataset 
$\mathcal{B}=\{\mathcal{B}_1,\mathcal{B}_2,\ldots,\mathcal{B}_M\}$ before transmission. 
The basic semantic unit is a triple, denoted by
\begin{equation}
    \varepsilon_i^j=(\psi_i,r_i^j,\phi_i),
    \label{eq:semantic_triple}
\end{equation}
where $\psi_i$ and $\phi_i$ denote the head and tail entities, respectively, and $r_i^j$ denotes the $j$-th relation associated with the entity pair $(\psi_i,\phi_i)$. 
The subscript $i$ identifies different entity pairs, while the superscript $j$ distinguishes different candidate relations for the same entity pair. 
Accordingly, the semantic set of UE $k$ can be written as $\mathcal{S}_k = \{\varepsilon_i^j \mid \varepsilon_i^j \text{ can be extracted from } \mathcal{D}_k \}$.

For each triple $\varepsilon_i^j$, let $\mathcal{N}_i^j$ denote the set of sample indices from which this triple can be extracted. 
Equivalently, the corresponding quadruple is given by
\begin{equation}
    \delta_i^j=(\psi_i,r_i^j,\phi_i,\mathcal{N}_i^j).
    \label{eq:quadruple}
\end{equation}
Based on \eqref{eq:quadruple}, the empirical probability of triple $\varepsilon_i^j$ associated with the entity pair $(\psi_i,\phi_i)$ is defined as
\begin{equation}
    \kappa(\varepsilon_i^j)
    =
    \frac{
    {\rm card}(\mathcal{N}_i^j)
    }{
    \sum_{l=1}^{J_i}
    {\rm card}(\mathcal{N}_i^l)
    },
    \label{eq:triple_probability}
\end{equation}
where ${\rm card}(\cdot)$ denotes the cardinality of a set, and $J_i$ represents the total number of candidate relations between $\psi_i$ and $\phi_i$. 
The probability in \eqref{eq:triple_probability} is normalized over the candidate relations associated with the same entity pair, satisfying
\begin{equation}
    \sum_{j=1}^{J_i}\kappa(\varepsilon_i^j)=1,
    \quad \forall i.
    \label{eq:local_probability_normalization}
\end{equation}
Therefore, $\kappa(\varepsilon_i^j)$ characterizes the empirical relation probability conditioned on the entity pair $(\psi_i,\phi_i)$, rather than a global probability over all triples in the KG. 
In the compression process, candidate relations are compared only within the same entity-pair-specific relation set, which avoids directly comparing raw occurrence frequencies across different entity pairs.

For semantic compression, relations with high empirical predictability can be omitted and recovered with the assistance of the shared PKG. 
Specifically, if the relation of a triple can be inferred from its entity pair and the probabilistic statistics stored in the PKG, the relation field is replaced by a lightweight omission token during transmission. 
The downgraded triple in the $n$-th compression round is represented as
\begin{equation}
    o_p^n=(\psi_p,\rho_n,\phi_p),
    \label{eq:downgraded_triple}
\end{equation}
where $\rho_n$ denotes the omission token and $p$ denotes the index of the downgraded triple. 
This operation reduces the semantic payload to be transmitted, but introduces additional computation for probabilistic inference and semantic reconstruction.

Rather than explicitly including all conditional probability matrices in the online optimization problem, the detailed PKG compression rules are absorbed into the compression strategy $\xi$ of $G_{\xi}(\cdot)$. 
Therefore, we define the \emph{Compression Ratio} (CR) to describe the proportion of downgraded triples, thereby linking the semantic compression process with the computation and communication costs. 
Let $u_k={\rm card}(\mathcal{U}_k)$ denote the number of triples in semantic component $\mathcal{U}_k$, where $k=0$ corresponds to the shared semantic component and $k\in\mathcal{K}$ corresponds to the private semantic component of UE $k$. 
The CR of semantic component $\mathcal{U}_k$ is defined as
\begin{equation}
    \Omega_k
    =
    \frac{
    \sum_{q=1}^{Q} d_{k,q}
    }{
    u_k
    },
    \quad \forall k\in\mathcal{K}\cup\{0\},
    \label{eq:cr_definition}
\end{equation}
where $d_{k,q}$ denotes the number of downgraded triples in the $q$-th compression round. 
The CR $\Omega_k$ provides a compact interface between the PKG-assisted semantic compression process and the subsequent UAV-RSMA optimization.

Based on the characteristics of semantic compression, the computation cost and communication load are modeled as functions of the CR. 
Specifically, the piecewise-linear computational cost $C(\Omega_k)$ is characterized through offline profiling of the PKG-assisted compression module. During the offline stage, the semantic compression algorithm is executed sequentially over multiple compression rounds. After each round, the resulting compression ratio and the corresponding computation cost are recorded, yielding a set of breakpoint pairs. The slope and intercept of each linear segment are then directly determined from two adjacent breakpoint pairs, resulting in a piecewise-linear approximation of the computation cost.

\begin{figure}[!h]
    \centering
    \includegraphics[width=0.9\columnwidth]{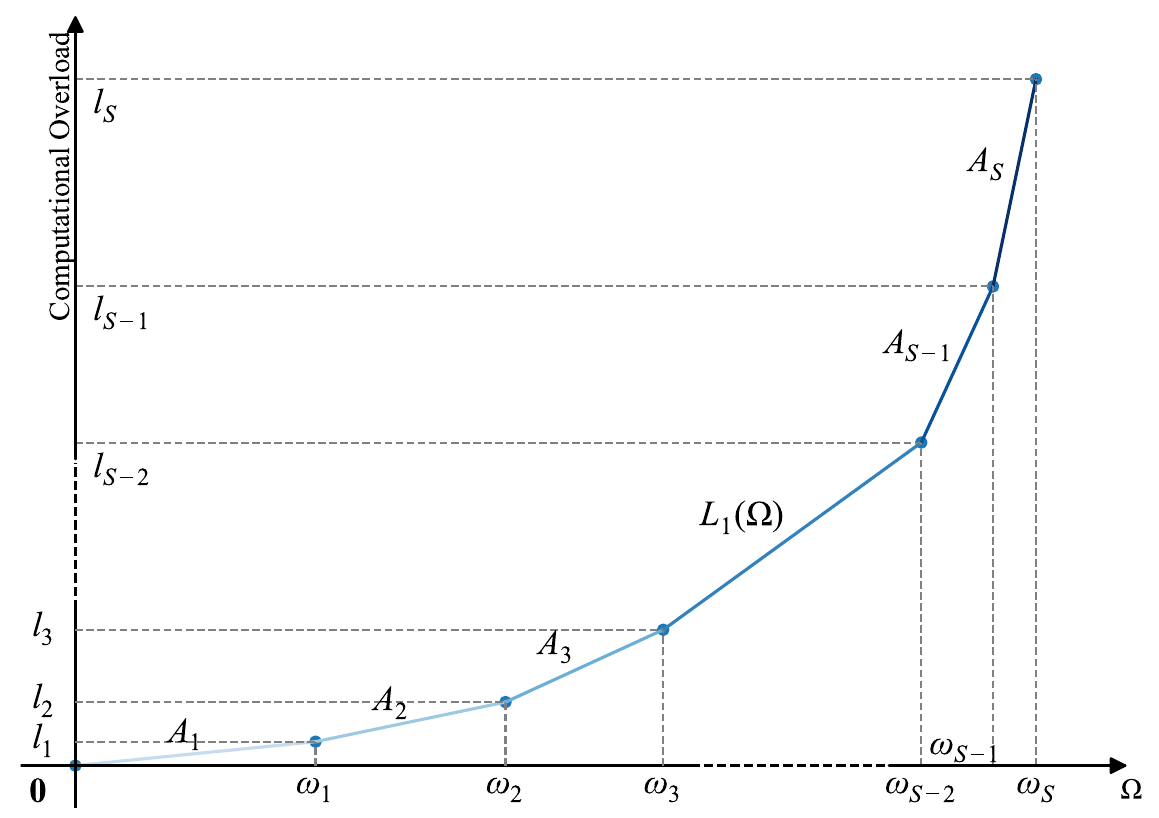}
    \caption{Illustration of the relationship between computational cost $C(\Omega)$ and semantic compression ratio $\Omega$.}
    \label{fig:CR_tradeoff}
\end{figure}

Hence, for the $s$-th segment, the computational cost in time slot $t$ can be expressed as
\begin{equation}
    C(\Omega_k[t]) = A_s \Omega_k[t] + B_s,
    \label{eq:computation_cost_cr}
\end{equation}
where $A_s$ and $B_s$ denote the slope and intercept corresponding to the CR interval of the $s$-th compression segment. 
The relationship between the computational cost and the CR is illustrated in Fig.~\ref{fig:CR_tradeoff}.

For communication overhead, equal-length coding is employed to encode the triples.
Specifically, let $L_e$ denote the fixed bit-length required to uniquely represent a entity in the knowledge base. For simplicity, both the head entity $\psi_i$ and the tail entity $\phi_i$ are encoded using $L_e$ bits. Since the relation $r_i^j$ typically contains richer semantic association information, it is represented using $2L_e$ bits. Therefore, a complete semantic triple occupies $4L_e$ bits before compression. Since the omission token $\rho_n$ can be represented using very few bits, it is neglected during transmission. Moreover, according to the semantic compression model, a compression ratio $\Omega$ corresponds to omitting semantic relations with an equivalent communication overhead reduction of $2L_e\Omega$ bits per triple. Consequently, the communication overhead after semantic compression in time slot $t$ can be expressed as

\begin{equation}
    D(\Omega[t]) = 4L_e - 2L_e\Omega[t], \quad \Omega[t]\in[0,1].
\end{equation}

Based on the explicit definitions of the shared and private compression ratios, the overall semantic compression vector for the system is given by
\begin{equation}
    \bm{\Omega} = [\Omega_0, \Omega_1, \dots, \Omega_K]^T.
\end{equation}

Subsequently, by jointly considering the semantic compression performed at the UAV and the downlink RSMA transmission, the total time delay for the private information of UE $k$ in time slot $t$ can be expressed as
\begin{equation}
    \tau_k[t] = \frac{C(\Omega_k[t])}{f_k} + \frac{u_k D(\Omega_k[t])}{R^p_k[t]}, \quad \forall k \in \mathcal{K}, \forall t \in \mathcal{T},
\end{equation}
where $f_k$ denotes the computation capability allocated for UE $k$. To focus on the interplay among semantic compression, resource allocation, and UAV trajectory design, the computation capability is assumed to remain constant during the mission. Dynamic computation frequency allocation is left as a future extension.
Similarly, the transmission delay for the shared information is given by
\begin{equation}
    \tau_0[t] = \frac{C(\Omega_0[t])}{f_0} + \frac{u_0 D(\Omega_0[t])}{R^s_0[t]}, \quad \forall t \in \mathcal{T},
\end{equation}
where $f_0$ denotes the computation capability allocated for compressing the shared information and is also assumed to be constant throughout the mission.

To evaluate semantic reconstruction reliability, the conventional triple-matching accuracy in~\cite{RSMA5} treats all semantic triples equally and can be expressed as
\begin{equation}
    \Gamma_k(\mathcal{S}_k,\mathcal{S}'_k)
    =
    \frac{
    \sum_{i=1}^{I}
    \min\{\Upsilon(\mathcal{S}_k,\varepsilon_i),
    \Upsilon(\mathcal{S}'_k,\varepsilon_i)\}
    }{
    \sum_{i=1}^{I}
    \Upsilon(\mathcal{S}_k,\varepsilon_i)
    },
    \label{eq:unweighted_semantic_accuracy}
\end{equation}
where $I$ is the number of distinct triples in $\mathcal{S}_k$, $\mathcal{S}'_k$ denotes the recovered semantic set of UE $k$, and $\Upsilon(\mathcal{S},\varepsilon_i)$ denotes the number of occurrences of triple $\varepsilon_i$ in $\mathcal{S}$.

However, different triples may have different importance in the PKG. Losing a PKG-important or task-critical triple may cause more severe semantic distortion than losing a less informative triple. Therefore, the unweighted metric in \eqref{eq:unweighted_semantic_accuracy} may overestimate the reconstruction quality when important semantic triples are missing. To better capture the loss of important semantic information, we introduce a PKG-weighted semantic accuracy metric as
\begin{equation}
    \widetilde{\Gamma}_k(\mathcal{S}_k,\mathcal{S}'_k)
    =
    \frac{
    \sum_{i=1}^{I}
    \kappa(\varepsilon_i)
    \min\{\Upsilon(\mathcal{S}_k,\varepsilon_i),
    \Upsilon(\mathcal{S}'_k,\varepsilon_i)\}
    }{
    \sum_{i=1}^{I}
    \kappa(\varepsilon_i)
    \Upsilon(\mathcal{S}_k,\varepsilon_i)
    },
    \label{eq:weighted_semantic_accuracy}
\end{equation}
where $\kappa(\varepsilon_i)$ denotes the PKG-derived probabilistic weight of triple $\varepsilon_i$. Compared with \eqref{eq:unweighted_semantic_accuracy}, \eqref{eq:weighted_semantic_accuracy} assigns larger weights to more informative triples, and thus can better reflect the impact of losing key semantic information.

\begin{figure}[t]
    \centering
    \includegraphics[width=0.85\columnwidth]{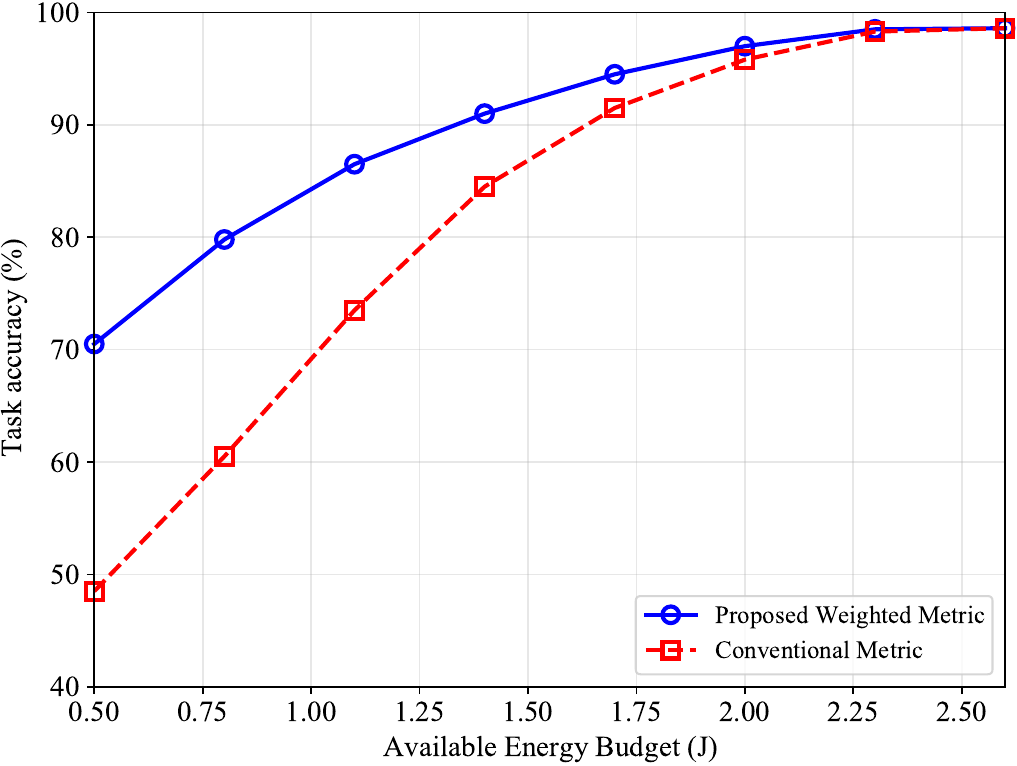}
    \caption{Task accuracy comparison under different available energy budgets.}
    \label{fig:metric}
\end{figure}

To further demonstrate the practical significance of the proposed weighted semantic metric, Fig.~\ref{fig:metric} compares the downstream task accuracy achieved by the weighted and conventional unweighted semantic metrics under different available resource budgets.

It can be observed that both metrics achieve comparable task completion rates when the available energy budget is sufficiently large, since adequate resources allow both important and non-critical semantic information to be preserved during transmission. However, under stringent energy constraints, the task completion rate associated with the unweighted metric decreases rapidly because it treats all semantic triples equally and may sacrifice task-critical triples to satisfy the semantic accuracy requirement. In contrast, the proposed weighted metric prioritizes PKG-important semantic triples and therefore maintains a substantially higher task completion rate. 
These results validate that the proposed metric provides a more faithful characterization of downstream semantic requirements and improves the robustness of semantic communications in resource-constrained wireless systems.

Since the recovered semantic set $\mathcal{S}'_k$ depends on the retained shared and private semantic components after compression, the weighted semantic accuracy can be further regarded as a function of the compression ratios,
\begin{equation}
    \widetilde{\Gamma}_k(\mathcal{S}_k,\mathcal{S}'_k) = \widetilde{\Gamma}_k\big(\Omega_0[t],\Omega_k[t];\boldsymbol{\kappa}\big).
    \label{eq:omega_dependent_accuracy}
\end{equation}
A larger compression ratio removes more semantic triples, and thus the retained semantic information generally decreases with $\Omega_0[t]$ and $\Omega_k[t]$. Therefore, $\widetilde{\Gamma}_k(\Omega_0[t],\Omega_k[t];\boldsymbol{\kappa})$ is generally non-increasing with respect to the semantic compression ratios.
\subsection{UAV Mobility and Channel Model}
Without loss of generality, the total UAV flight time $T_v$ is discretized into $T$ equal-length time slots, each of duration $\delta$. This division yields a set of time slots represented by $\mathcal{T} = \{1, 2, \ldots, T\}$, where $T_v = \delta T$. 

We assume that the UAV departs from a specified initial position within a finite flight duration $T$ to support the restoration of communication networks in the target area.
The maximum flight speed of the UAV is limited to $V_{\text{max}}$, thus the maximum movement distance of the UAV within each time slot is $\delta V_{\text{max}}$. 
Given the relatively minute displacement of the UAV within each time slot, we reasonably assume that the distance between the UAV and each UE remains constant within a time slot.
To simplify the problem and effectively avoid collisions, the flight altitude of the UAV is fixed at $H$. The takeoff and landing phases are neglected.
Considering a three-dimensional Cartesian coordinate system, the position of the UAV on the horizontal plane at any time point $t \in [0, T]$ is represented as $\mathbf{q}[t] = (x[t], y[t])$. Hence, we have
\begin{equation}
    ||\mathbf{q}[t+1] - \mathbf{q}[t]|| \leq (\delta V_{\text{max}})^2, \quad \forall t,
    \label{distance1}
\end{equation}
where $||\cdot||$ denotes Euclidean norm. Additionally, each UE $k \in \mathcal{K}$ is randomly positioned at fixed locations $L_i = (x_i, y_i)$ and $L_j = (x_j, y_j)$ respectively.
Consequently, the distances between the UAV and UEs in each time slot $t$ can be expressed as
\begin{equation}
    d_k[t] = \sqrt{ H^2 + ||\mathbf{q}[t] - L_k||^2}, \quad \forall k, t.
\end{equation}

In temporary communication setups, radio frequency (RF) channels are often employed to provide wide-area coverage, penetrate obstacles, and adapt to complex environments. According to 3rd generation partnership project (3GPP) specifications, the RF channel between the UAV and UEs is modeled as a probabilistic LoS channel~\cite{channel}.

Based on previous work~\cite{RF}, we consider a quasi-static wireless channel, where the channel condition is assumed to remain constant within each time slot. To reflect more realistic propagation characteristics, we incorporate both shadowing effects and probabilistic LoS conditions.
The path loss $\mu_k$ (in decibels) between the UAV and the $k$-th UE in time slot $t$ is defined as
\begin{equation}
    \mu_k[t] = \begin{cases} 
        20 \log_{10} d_k[t] + \zeta^{\text{LoS}} + X_{\sigma}^{\text{LoS}}, & \text{if LoS link}, \\
        20 \log_{10} d_k[t] + \zeta^{\text{NLoS}} + X_{\sigma}^{\text{NLoS}}, & \text{if NLoS link},
    \end{cases}
    \label{pathloss}
\end{equation}
where $d_i[t]$ represents the distance between the UAV and the $k$-th UE in time slot $t$,
$\zeta^{\text{LoS}}$ and $\zeta^{\text{NLoS}}$ represent the average path loss values for LoS and NLoS conditions, $X_{\sigma}^{\text{LoS}}$ and $X_{\sigma}^{\text{NLoS}}$ represent shadowing effects, respectively.
The LoS/NLoS link between the UAV and the $k$-th UE is randomly determined based on the LoS/NLoS probabilities defined as
\begin{equation}
    P_k^{\text{LoS}}[t] = \frac{1}{{1 + \eta {e^{ - \gamma ({\varphi _k[t]} - \eta )}}}},
    \label{PLoS}
\end{equation}
\begin{equation}
    P_k^{\text{NLoS}}[t] = 1 - P_k^{\text{LoS}}({\varphi _k[t]}),
    \label{PNLoS}
\end{equation}
where $\eta$ and $\gamma$ are two environmental parameters related to the deployment model, $\varphi_k[t]$ represents the elevation angle between the UAV and the $k$-th UE in time slot $t$, which is calculated as
\begin{equation}
    {\varphi _k[t]} = \frac{180^{\circ}}{\pi}{\arcsin }\left( {\frac{H}{{{d_k[t]}}}} \right),
    \label{angle}
\end{equation}
where $H$ is the flight altitude of the UAV. Consequently, the average path loss in dB between the UAV and UE $k$ is
\begin{equation}
    h_k[t] = \mu_k^{\text{LoS}}[t]P_k^{\text{LoS}}[t] + \mu_k^{\text{NLoS}}[t]P_k^{\text{NLoS}}[t],
    \label{averagepathloss}
\end{equation}
where $h_k[t]$ represents the average path loss expressed in dB. In the signal model, its corresponding linear-scale representation is employed to characterize the effective channel vector $\mathbf{h}_k[t]\in\mathbb{C}^{M\times1}$.

\subsection{RSMA Transmission Model}
In each time slot $t \in \mathcal{T}$, the UAV serves as an aerial BS to transmit both the shared and private information to the UEs using downlink RSMA. The transmitted signal of the UAV can be expressed as
\begin{equation}
    \mathbf{x}[t] = 
    \underbrace{\sqrt{p_0[t]}\mathbf{w}_0 m_0[t]}_{\text{shared information}} + 
    \underbrace{\sum_{k=1}^{K} \sqrt{p_k[t]}\mathbf{w}_k m_k[t]}_{\text{private information}},
\end{equation}
where $\mathbf{w}_0$ and $\mathbf{w}_k$ denote the beamforming vectors for the shared and private information, respectively. The transmission powers allocated to these components are $p_0[t]$ and $p_k[t]$ in time slot $t$.
The received signal at UE $k$ during time slot $t$ is given by
\begin{equation}
    \mathbf{y}_k[t] = \mathbf{h}_k^H[t]\mathbf{x}[t] + n_k[t],
\end{equation}
where $\mathbf{h}^H_k[t]$ denotes the channel vector between the UAV and UE $k$ in time slot $t$, and $n_k \sim \mathcal{CN}(0, \sigma_n^2)$ represents the AWGN with zero mean and variance $\sigma_n^2$.

The achievable rate for UE $k$ to decode the shared information is expressed as
\begin{equation}
    R^s_k[t] = B \log_2 \!\left( 1 + 
    \frac{p_0[t]|\mathbf{h}_k^H[t]\mathbf{w}_0|^2}{
    \sum_{j=1}^{K} p_j[t]|\mathbf{h}_k^H[t]\mathbf{w}_j|^2 + B\sigma^2} \right),
\end{equation}
where $B$ denotes the system bandwidth.  
To ensure successful decoding of the shared information at all users,  the shared transmission rate is given by~\cite{RSMA1}
\begin{equation}
    R^s_0[t] = \min_{k \in \mathcal{K}} R^s_k[t].
\end{equation}
After the shared information is decoded and subtracted from the received signal, UE $k$ proceeds to decode its private information. The achievable private rate is thus given by
\begin{equation}
    R^p_k[t] = B \log_2 \!\left( 1 + 
    \frac{p_k[t]|\mathbf{h}_k^H[t]\mathbf{w}_k|^2}{
    \sum_{j=1, j \neq k}^{K} p_j[t]|\mathbf{h}_k^H[t]\mathbf{w}_j|^2 + B\sigma^2} \right).
\end{equation}

According to the coding rules of RSMA strategy~\cite{RSMA1}, at receiver side, users first decode the received information with SIC technique and obtain the received semantic information.

\section{Problem Formulation and Algorithm}
In this section, we first formulate the problem to minimize the system energy consumption comprehensively considering the communication, computation and flight during transmission. Then, we propose an alternative algorithm with optimizing subproblems of joint power allocation and beamforming design, semantic compression ratio, and UAV trajectory optimization.
\subsection{Problem Formulation}
We consider the total energy consumption of the UAV networks, which consists of three parts: computation energy, communication energy, and UAV flight energy.

\paragraph{Computation Energy Consumption}
The total computation energy includes the energy consumed for compressing both the shared and private semantic information. Since the CR $\Omega_k[t]$ determines the proportion of semantic relations retained during the PKG-assisted semantic compression process, it directly dictates the computational workload $C(\Omega_k[t])$ required for semantic processing. The total computation energy can be expressed as
\begin{equation}
    E_1[t] = \sum_{k=0}^K \lambda C(\Omega_k[t]) f_k^2, \quad \forall t \in \mathcal{T}, 
\end{equation}
where $\lambda$ is a constant coefficient representing the effective switched capacitance.

\paragraph{Communication Energy Consumption} For the communication part, the total energy consumption includes both shared and private transmissions. Since the CR $\Omega_k[t]$ determines the amount of semantic payload $D(\Omega_k[t])$ to be transmitted, it directly affects the communication energy consumption through the transmission duration. The total communication energy can be expressed as
\begin{equation}
    E_2[t] = p_0[t]\frac{u_0 D(\Omega_0[t])}{R^s_0[t]} + \sum_{k=1}^{K} p_k[t]\frac{u_k D(\Omega_k[t])}{R^p_k[t]},
    \quad \forall t \in \mathcal{T},
    \label{eq:E_comm}
\end{equation}

\paragraph{UAV Flight Energy Consumption}
The propulsion power consumption follows the widely adopted rotary-wing UAV model in~\cite{Challenge2}, whose detailed expression is omitted for brevity. The corresponding aerodynamic parameters are listed in Table~II.
The UAV velocity at time slot $t$ is given by
\begin{equation}
    \bm{\nu}[t]=\frac{\mathbf q[t+1]-\mathbf q[t]}{\delta}, \quad \forall t\in\mathcal T.
\end{equation}

Accordingly, the flight energy consumption in time slot $t$ is
\begin{equation}
    E_3[t]=\delta P(\|\bm{\nu}[t]\|), \quad \forall t\in\mathcal T.
\end{equation}

Unlike conventional UAV-enabled RSMA optimization problems where the transmission payload is fixed, the PKG-assisted semantic communication model introduces semantic partitioning and probabilistic semantic compression into the optimization process. Notably, rather than relying on black-box non-linear curve fitting typical in conventional deep learning-based semantic models, the computational and communication costs in our framework are explicitly derived from the profiling of the PKG relation omission process. Consequently, the amount of information to be transmitted becomes a function of the semantic compression ratios and the semantic structure extracted from the PKG. This establishes an additional coupling between semantic processing and communication resource allocation, making the optimization variables jointly affect both semantic representation and transmission efficiency.
By integrating all components of energy consumption, the overall optimization problem is formulated as follows. 
A weight factor $\omega$ is introduced to balance the different magnitudes of UAV propulsion energy and communication/computation energy components in the objective function. In practical UAV systems, propulsion energy is typically much higher than communication and computation energy consumption. Therefore, $\omega$ acts as a scale-normalization factor to prevent the optimization objective from being dominated by the flight energy term and to maintain numerically comparable contributions among heterogeneous energy components,

\begin{subequations}\label{eq:p1}
\begin{align}
    \min_{\mathbf{p}, \mathbf{W}, \boldsymbol{\Omega}, \mathbf{Q}} \quad & E = \sum_{t=0}^T(E_1[t] + E_2[t] + \omega E_3[t]),  \label{p1-main} \tag{39}\\
    \text{s.t.} \quad & \sum_{k=0}^{K} p_k[t] \leq P_{\text{max}}, \quad \forall t, \label{p1-power1} \\
    & \tau_k[t] \leq \delta, \quad \forall k \in \mathcal{K} \cup \{0\}, t, \label{p1-time}\\
    & \widetilde{\Gamma}_k\big(\Omega_0[t],\Omega_k[t];\boldsymbol{\kappa}\big) \geq \widetilde{\Gamma}_{\min}, \quad \forall k \in \mathcal{K}, t, \label{p1-accuracy}\\
    & p_k[t] \geq 0, \quad \forall k \in \mathcal{K} \cup \{0\}, t, \label{p1-power2}\\
    & 0 \leq \Omega_k[t] \leq 1, \quad \forall k \in \mathcal{K} \cup \{0\}, t, \label{p1-CR}\\
    & \|\mathbf{w}_k[t]\| = 1, \quad \forall k \in \mathcal{K} \cup \{0\}, t, \label{p1-beamforming}\\
    & |\mathbf{q}[t]-\mathbf{q}[t-1]|\leq V_{\text{max}}\delta. \label{p1-trajectory}
\end{align}
\end{subequations}

The optimization problem \eqref{eq:p1} minimizes the total energy consumption $E$ over all time slots, subject to several system constraints. 
Specifically, constraint \eqref{p1-power1} limits the total transmit power at each time slot to the maximum available power $P_{\text{max}}$. Constraint \eqref{p1-time} ensures that the transmission delay of each user does not exceed the maximum tolerable delay $\delta$. 
Constraint \eqref{p1-accuracy} guarantees that the semantic accuracy of all users remains above the minimum threshold $\widetilde{\Gamma}_{\text{min}}$.
Constraint \eqref{p1-power2} enforces the non-negativity of transmit power.
Constraint \eqref{p1-CR} confines the semantic compression ratio to the range $[0,1]$.
The unit-norm requirement of beamforming vectors is imposed by constraint \eqref{p1-beamforming}.
Constraint \eqref{p1-trajectory} limits the UAV’s displacement within each time slot, ensuring that its movement does not exceed the maximum flight distance $V_{\text{max}}\delta$.

To solve this non-convex problem, we employ the alternating optimization method, which decomposes the original problem into three subproblems for iterative optimization.
\subsection{Joint Optimization of Power Allocation and Beamforming Design}
When the semantic compression rate $\boldsymbol{\Omega}$ and UAV trajectory $\mathbf{Q}$ are given, the constraints \eqref{p1-accuracy}, \eqref{p1-CR}, and \eqref{p1-trajectory} can be omitted. Accordingly, $E_1$ and $E_3$ become constants. To simplify the optimization, we define $\mathbf{v}_k[t]=\sqrt{p_k[t]}\mathbf{w}_k[t]$, such that the actual transmit power and received signal power are expressed as $\|\mathbf{v}_k[t]\|^2$ and $|\mathbf{h}_k^H[t]\mathbf{v}_k[t]|^2$, respectively. We further introduce a relaxation parameter $\boldsymbol{\zeta}[t] = [\zeta_0[t], \zeta_1[t], \dots, \zeta_K[t]]^T$, and problem (\ref{eq:p1}) can be equivalently transformed into
\allowdisplaybreaks  
\begin{subequations}\label{eq:p2}
\begin{align}
    \min_{\mathbf{V}, \boldsymbol{\zeta}} \quad & E = 2L_e \sum_{t=0}^T\biggl[\frac{u_0[t](2 - \Omega_0[t]){\|\mathbf{v}_0[t]\|^2}}{\zeta_0[t]} \nonumber \\
    &\quad + \sum_{j=1}^K\frac{u_j[t](2 - \Omega_j[t]){\|\mathbf{v}_j[t]\|^2}}{\zeta_j[t]}\biggl] \label{p2-main} \tag{40}\\
    \text{s.t.} \quad & \sum_{k=0}^{K} {\|\mathbf{v}_k[t]\|^2} \leq P_{\text{max}}, \quad \forall t \label{p2-power}\\
    & \zeta_k[t] \geq 0, \quad \forall k \in \mathcal{K} \cup \{0\}, t, \\
    & \frac{2L_{e}u_k[t](2 - \Omega_k[t])}{\zeta_k[t]} \leq \delta_k[t], \quad \forall k \in \mathcal{K}\cup\{0\}, t, \label{p2-time} \\
    &\zeta_0[t] \leq B \log_2 \left( 1 + \frac{{|\mathbf{h}_k^H[t] \mathbf{v}_0[t]|^2}}{\sum_{j=1}^K {|\mathbf{h}_k^H[t] \mathbf{v}_j[t]|^2} + B\sigma^2} \right), \nonumber \\
    & \forall k \in \mathcal{K}, t, \label{p2-trans1} \\
    &\zeta_k[t] \leq B \log_2 \left( 1 + \frac{{|\mathbf{h}_k^H[t] \mathbf{v}_k[t]|^2}}{\sum_{j=1, j \neq k}^K {|\mathbf{h}_k^H[t] \mathbf{v}_j[t]|^2} + B\sigma^2} \right), \nonumber \\
    & \forall k \in \mathcal{K}, t, \label{p2-trans2}
\end{align}
\end{subequations}
where $\delta_k[t]$ in constraint \eqref{p2-time} is a constant defined as $\delta_k[t] = \delta - (A_k[t] \Omega_k[t] + B_k[t])/f_k$.

It should be noted that the auxiliary variables $\boldsymbol{\zeta}[t]$ serve as lower bounds on the achievable transmission rates. Since the perspective function $\|\mathbf v_k[t]\|^2/\zeta_k[t]$ is strictly decreasing with respect to $\zeta_k[t]>0$ and jointly convex with respect to $\|\mathbf v_k[t]\|$ and $\zeta_k[t]$~\cite{boyd_convex}, the optimal solution always drives $\zeta_k[t]$ to its largest feasible value. Consequently, constraints \eqref{p2-trans1} and \eqref{p2-trans2} must be active at the optimum. Therefore, introducing $\boldsymbol{\zeta}[t]$ does not cause any relaxation gap, guaranteeing the equivalence between Problems~\eqref{eq:p1} and~\eqref{eq:p2}. Furthermore, the objective function and constraint \eqref{p2-time} are inherently convex. The remaining non-convexity stems solely from the SINR constraints \eqref{p2-trans1} and \eqref{p2-trans2}, which we sequentially convexify using the SCA technique.

To convexify the SINR constraints in \eqref{p2-trans1} and \eqref{p2-trans2},
auxiliary variables $\boldsymbol{\alpha}[t]$, $\boldsymbol{\beta}[t]$, $\boldsymbol{\gamma}[t]$, and $\boldsymbol{\eta}[t]$ are introduced, where $\boldsymbol{\alpha}[t]$ and $\boldsymbol{\beta}[t]$ represent the SINR variables,
while $\boldsymbol{\gamma}[t]$ and $\boldsymbol{\eta}[t]$ are introduced to decouple the denominator terms. Accordingly,
\begin{equation}
    \alpha_k[t] \leq \frac{|\mathbf{h}_k^H[t]\mathbf{v}_0[t]|^2}{\gamma_k[t]} \leq \frac{|\mathbf{h}_k^H[t]\mathbf{v}_0[t]|^2}{\sum_{j=1}^K |\mathbf{h}_k^H[t]\mathbf{v}_j[t]|^2 + B\sigma^2},
\end{equation}
\begin{equation}
    \beta_k[t] \leq \frac{|\mathbf{h}_k^H[t]\mathbf{v}_k[t]|^2}{\eta_k[t]} \leq \frac{|\mathbf{h}_k^H[t]\mathbf{v}_k[t]|^2}{\sum_{j=1, j \neq k}^K |\mathbf{h}_k^H[t]\mathbf{v}_j[t]|^2 + B\sigma^2}.
\end{equation}
\begin{equation}
    \gamma_k[t] \geq \sum_{j=1}^{K}|\mathbf{h}_k^H[t] \mathbf{v}_j[t]|^2 + B\sigma^2  , \quad \forall k \in \mathcal{K}, t, \\
    \label{p4-transa-1}
\end{equation}
\begin{equation} 
    \alpha_k[t] \gamma_k[t] \leq |\mathbf{h}_k^H[t] \mathbf{v}_0[t]|^2, \quad \forall k \in \mathcal{K}, t,
    \label{p4-transa-2}
\end{equation}
\begin{equation}
    \eta_k[t] \geq \sum_{j=1,j \neq k}^{K} |\mathbf{h}_k^H[t] \mathbf{v}_j[t]|^2 + B\sigma^2, \quad \forall k \in \mathcal{K}, t, \\
    \label{p4-transb-1}
\end{equation}
\begin{equation} 
    \beta_k[t] \eta_k[t] \leq |\mathbf{h}_k^H[t] \mathbf{v}_k[t
    ]|^2, \quad \forall k \in \mathcal{K}, t.
    \label{p4-transb-2}
\end{equation}
where \eqref{p4-transa-1}) and \eqref{p4-transb-1} are convex,
while constraint \eqref{p4-transa-2} remains non-convex.
It can be rewritten as
\begin{equation}
(\alpha_{k}[t]+\gamma_{k}[t])^{2}
+\Big(-(\alpha_{k}[t]-\gamma_{k}[t])^{2}\Big) + \Big(-4|\mathbf{h}_{k}^{H}[t]\mathbf{v}_{0}[t]|^{2}\Big)
\leq 0,
\end{equation}
which is in the standard difference-of-convex (DC) form. To convexify it, the two concave terms are linearized via the first-order Taylor approximation,
\allowdisplaybreaks
\begin{align}
    &(\alpha_{k}[t]+\gamma_{k}[t])^{2}  \nonumber\\
    & - (\alpha_k^{(n)}[t] - \gamma_k^{(n)}[t])^2
    - 2(\alpha_k^{(n)}[t] - \gamma_k^{(n)}[t])\big(\alpha_k[t]-\alpha_k^{(n)}[t]\big) \nonumber\\
    & + 2(\alpha_k^{(n)}[t] - \gamma_k^{(n)}[t])\big(\gamma_k[t]-\gamma_k^{(n)}[t]\big) \nonumber\\
    & - 4\big|\mathbf{h}_{k}^{H}[t]\mathbf{v}_{0}^{(n)}[t]\big|^{2} \nonumber\\
    & - 8\,\mathbb{R}\mathrm{e}\!\left\{
        \left(\mathbf{v}_0^{(n)}[t]\right)^{H}
        \mathbf{h}_k[t]\mathbf{h}_k^{H}[t]
        \big(\mathbf{v}_0[t]-\mathbf{v}_0^{(n)}[t]\big)
    \right\} \leq 0,
    \label{p4-transa-2conv}
\end{align}
where the superscript $(n)$ denotes the value at the $n$-th SCA iteration, which yields a convex constraint after the first-order Taylor approximation.

Similarly, constraint (\ref{p4-transb-2}) can be convexified via the first-order Taylor approximation
\allowdisplaybreaks
\begin{align}
    &(\beta_{k}[t]+\eta_{k}[t])^{2} \nonumber \\
    & - (\beta_k^{(n)}[t] - \eta_k^{(n)}[t])^2 
         - 2 (\beta_k^{(n)}[t] - \eta_k^{(n)}[t]) \big(\beta_k[t] - \beta_k^{(n)}[t]\big) \nonumber \\
    & + 2 (\beta_k^{(n)}[t] - \eta_k^{(n)}[t]) \big(\eta_k[t] - \eta_k^{(n)}[t]\big) \nonumber \\
    & -4\,\bigl\lvert\mathbf{h}_{k}^{H}[t]\mathbf{v}_{0}^{(n)}[t]\bigr\rvert^{2} \nonumber \\
    & -8\,\mathbb{R}\mathrm{e}\!\left\{\left( \mathbf{v}_0^{(n)}[t] \right)^H \mathbf{h}_k[t] \mathbf{h}_k^H[t]\big(\mathbf{v}_0[t]-\mathbf{v}_0^{(n)}[t]\big)\right\} \leq 0,
    \label{p4-transb-2conv}
\end{align}

Therefore, problem \eqref{eq:p2} can be further approximated as
\begin{subequations}\label{eq:p5}
\begin{align}
    \min_{\substack{\mathbf{V}, \boldsymbol{\zeta}, \boldsymbol{\alpha}, \\
        \boldsymbol{\beta}, \boldsymbol{\gamma}, \boldsymbol{\eta}}}
    \quad & E = 2L_{e} \sum_{t=0}^T \biggl[ \frac{u_0[t](2 - \Omega_0[t]) \|\mathbf{v}_0[t]\|^2}{\zeta_0[t]} \nonumber \\
    &\quad + \sum_{j=1}^K \frac{u_j[t](2 - \Omega_j[t]) \|\mathbf{v}_j[t]\|^2}{\zeta_j[t]} \biggr], \tag{50} \\
    \text{s.t.} \quad & \zeta_k[t], \eta_k[t], \gamma_k[t] \geq 0, \quad \forall k \in \mathcal{K} \cup \{0\}, t, \\
    &\zeta_0[t] \leq B \log_2 (1 + \alpha_k[t] ), \quad \forall k \in \mathcal{K}, t \\
    &\zeta_k[t] \leq B \log_2 (1 + \beta_k[t]), \quad \forall k \in \mathcal{K}, t \\
    & \eqref{p2-power},\; \eqref{p2-time},\; \eqref{p4-transa-1},\; \eqref{p4-transb-1},\; \eqref{p4-transa-2conv},\; \eqref{p4-transb-2conv},
\end{align}
\end{subequations}
which can be solved by existing optimization tool box.

\subsection{Semantic Compression Ratio Optimization}
When the power allocation $\mathbf{P}$, beamforming $\mathbf{W}$, and UAV trajectory $\mathbf{Q}$ are given, the constraints (\ref{p1-power1}) (\ref{p1-power2}), (\ref{p1-beamforming}) and (\ref{p1-trajectory}) can be omitted. Accordingly, $E_1$ becomes constant. Since the semantic accuracy constraint is monotonically related to the compression ratios, it can be transformed into the following conservative compression-ratio bounds
\begin{equation}
    0\leq \Omega_0[t]\leq \Theta_0,\quad
    0\leq \Omega_k[t]\leq \Theta_k,
    \quad \forall k\in\mathcal K,t,
    \label{eq:Theta_bound}
\end{equation}
where $\Theta_0$ and $\Theta_k$ are obtained through offline semantic reconstruction tests under the target accuracy threshold $\widetilde{\Gamma}_{\min}$. 

As shown in~Fig. \ref{fig:CR_tradeoff}, the slopes of $C(\Omega)$
do not decrease, so $C(\Omega)$ is convex and can be represented as
\begin{equation}
    C(\Omega_k[t]) = \max_{i=1,2,3\dots,S} A_i[t]\Omega_k[t] + B_i[t].
\end{equation}
So the compression rate optimization subproblem can be written as
\allowdisplaybreaks  
\begin{subequations}\label{eq:p7}
\begin{align}
    \min_{\boldsymbol{\Omega}} \quad & E = \sum_{t=0}^T \Bigg( \lambda \sum_{k=0}^K \Big(\max\limits_{i=1,2,3\dots,S} A_i[t]\Omega_k[t] + B_i[t]\Big) f_k^2 \nonumber \\
    &\quad + 2L_{e} \biggl[ \frac{u_0[t](2 - \Omega_0[t]) p_0[t]}{R^s_0[t]} \nonumber\\
    &\quad + \sum_{j=1}^K \frac{u_j[t](2 - \Omega_k[t]) p_j[t]}{R^p_j[t]} \biggr] \Bigg) \tag{53}\\
    \text{s.t.} \quad & \frac{\max\limits_{i=1,2,3\dots,S} \Big(A_i[t]\Omega_k[t] + B_i[t]\Big)}{f_i} \nonumber\\
    \quad & + \frac{2L_{e} (2 - \Omega_k[t]) u_k[t]}{R^p_k[t]} \leq \delta_k, \quad \forall k \in \mathcal{K}, t, \label{p6-time1} \\
    \quad & \frac{\max\limits_{i=1,2,3\dots,S} \Big(A_i[t]\Omega_0[t]+ B_i[t]\Big)}{f_0} \nonumber \\
    \quad & + \frac{2L_{e} (2 - \Omega_0[t]) u_0[t]}{R^s_0[t]} \leq \delta_0, \quad \forall t, \label{p6-time2} \\
    & 0 \leq \Omega_k[t] \leq \Theta_k, \quad \forall k \in \mathcal{K} \cup \{0\}, t. \label{p6-CR}
\end{align}
\end{subequations}

It can be observed that the objective function in problem (\ref{eq:p7}) is convex, and the constraints are obvious convex constraints. Therefore, existing optimization tools can be utilized to solve the problem.

\subsection{UAV Trajectory Optimization}
When the power allocation $\mathbf{P}$, beamforming $\mathbf{W}$, and semantic compression rate $\mathbf{\Omega}$ are given, the constraints \eqref{p1-power1}, \eqref{p1-power2}, \eqref{p1-accuracy} and \eqref{p1-beamforming} can be omitted. Accordingly, $E_1$ becomes constant.
Similarly, we introduce relaxation parameters $\boldsymbol{\zeta}[t] = [\zeta_0[t], \zeta_1[t], \dots, \zeta_K[t]]^T$, $\boldsymbol{\alpha}[t] = [\alpha_1[t], \dots, \alpha_K[t]]^T$ and $\boldsymbol{\beta}[t] = [\beta_1[t], \dots, \beta_K[t]]^T$, \eqref{eq:p1} can be equivalently transformed into

\allowdisplaybreaks  
\begin{subequations}\label{eq:p9}
\begin{align}
    \min_{\mathbf{Q}, \boldsymbol{\zeta}, \boldsymbol{\alpha}, \boldsymbol{\beta}} \quad & E = \sum_{t=0}^T \Bigg(2L_{e} \Biggl[ \frac{u_0[t](2 - \Omega_0[t]) p_0[t]}{\zeta_0[t]}  \nonumber \\ 
    & \quad + \sum_{j=1}^K \frac{u_j[t](2 - \Omega_j[t]) p_j[t]}{\zeta_j[t]} \Biggr] \nonumber + w \delta P ( \bigl\| \bm{\nu}[t] \bigr\| ) \Bigg), \tag{54} \\
    \text{s.t.} \quad & \zeta_k[t] \geq 0, \quad \forall k \in \mathcal{K} \cup \{0\}, t, \label{p9-trans}\\
    & \frac{2L_{e} u_k[t](2 - \Omega_k[t])}{\zeta_k[t]} \leq {\delta_k[t]} , \quad \forall k \in \mathcal{K} \cup \{0\}, t, \label{p9-time}\\
    &\zeta_0[t] \leq B \log_2 (1 + \alpha_k[t]), \quad \forall t \label{p9-trans1}\\
    &\zeta_k[t] \leq B \log_2 (1 + \beta_k[t]), \quad \forall k \in \mathcal{K}, t \label{p9-trans2}\\
    & \alpha_k[t] \leq \frac{p_0 |\mathbf{h}_k^H[t] \mathbf{w}_0[t]|^2}{\sum_{j=1}^K p_j |\mathbf{h}_k^H[t] \mathbf{w}_j[t]|^2 + B\sigma^2}, \nonumber \\
    & \forall k \in \mathcal{K}, t, \label{p9-transa}\\
    & \beta_k[t] \leq \frac{p_k |\mathbf{h}_k^H[t] \mathbf{w}_k[t]|^2}{\sum_{j=1, j \neq k}^K p_j |\mathbf{h}_k^H[t] \mathbf{w}_j[t]|^2 + B\sigma^2}, \nonumber \\
    & \forall k \in \mathcal{K}, t. \label{p9-transb}
\end{align}
\end{subequations}
To convexify the SINR constraints \eqref{p9-transa} and \eqref{p9-transb}, auxiliary variables $\boldsymbol{\gamma}[t]$ and $\boldsymbol{\eta}[t]$ are introduced to decouple the multi-user interference and noise components. Specifically, \eqref{p9-transa} and \eqref{p9-transb} are equivalently reformulated as
\begin{align}
    &\sum_{j=1}^{K} p_j[t] |\mathbf{h}_k^H[t] \mathbf{w}_j[t]|^2 + B\sigma^2 \leq \gamma_k[t], \quad \forall k, t, \label{p9-transa-1} \\
    &\alpha_k[t] \gamma_k[t] \leq p_0[t] |\mathbf{h}_k^H[t] \mathbf{w}_0[t]|^2, \quad \forall k, t, \label{p9-transa-2} \\
    &\sum_{j=1, j \neq k}^{K} p_j[t] |\mathbf{h}_k^H[t] \mathbf{w}_j[t]|^2 + B\sigma^2 \leq \eta_k[t], \quad \forall k, t, \label{p9-transb-1} \\
    &\beta_k[t] \eta_k[t] \leq p_k[t] |\mathbf{h}_k^H[t] \mathbf{w}_k[t]|^2, \quad \forall k, t. \label{p9-transb-2}
\end{align}
For tractability, the effective channel vector derived from the probabilistic LoS/NLoS channel model is approximated by its corresponding average large-scale fading component during the trajectory optimization process. Accordingly, since the beamforming vector is fixed, we have
\begin{equation}
h_k(\mathbf q[t]) = |\mathbf{h}_k^H[t] \mathbf{w}_0[t]| = \frac{\vartheta_0 G}{|\mathbf{q}[t]-\mathbf{L}_k|^2 + H^2},
\label{p9-transa-beaming}
\end{equation}
where $\mathbf{L}_k$ denotes the location of UE $k$, and $G$ is a constant related to the beam or array direction. The function $h_k(\mathbf q[t])$ is differentiable with respect to $\mathbf q[t]$. Applying the first-order Taylor approximation to the right-hand sides of the above bilinear constraints yields
\begin{align}
\alpha_k[t]\gamma_k[t] &\le p_0[t]h_k^{lb}(\mathbf q[t]), \label{flight1}\\
\beta_k[t]\eta_k[t] &\le p_k[t]h_k^{lb}(\mathbf q[t]). \label{flight2}
\end{align}

The propulsion power term $P(\|\bm{\nu}[t]\|)$ follows the rotary-wing UAV model in~\cite{Challenge2}. Since only the induced-power component is non-convex, we introduce a relaxation variable
$\nu[t]\ge\|\bm{\nu}[t]\|$.
Furthermore, an auxiliary variable $g[t]$ is introduced to represent the square-root term in the induced-power component, satisfying

\begin{equation}
g^2[t]\ge
\sqrt{1+\frac{\nu^4[t]}{4\nu_0^4}}
-\frac{\nu^2[t]}{2\nu_0^2}.
\end{equation}
This constraint can be equivalently transformed into

\begin{equation}
\frac{1}{g^2[t]}
\le
g^2[t]+\frac{\nu^2[t]}{\nu_0^2}.
\end{equation}
For any given local point ${\nu^{(n)}[t], g^{(n)}[t]}$ (where the superscript $(n)$ denotes the $n$-th iteration),
the right-hand side of (64) can be lower-bounded via first-order Taylor approximation,
since it is jointly convex with respect to $\nu[t]$ and $g[t]$~\cite{boyd_convex}.
Let $\chi^{lb}[t]$ denote this lower-bound function, which is given by
\begin{align}
\chi^{lb}[t] & \triangleq (g^{(n)}[t])^2 + 2g^{(n)}[t](g[t] - g^{(n)}[t]) \nonumber\\
&+ \frac{(\nu^{(n)}[t])^2}{\nu_0^2}  + \frac{2\nu^{(n)}[t]}{\nu_0^2}(\nu[t] - \nu^{(n)}[t]),  \forall t.
\end{align}

Accordingly, the propulsion power can be approximated by the following convex upper-bound
\begin{equation}
\widetilde{P}(\nu[t]) = P_0 \left( 1 + \frac{3\nu^2[t]}{U_{tip}^2} \right) + P_i g[t] + \frac{1}{2} d_0 \rho s A \nu^3[t].
\label{flight3}
\end{equation}
Therefore, problem (\ref{eq:p9}) can be further approximated as
\allowdisplaybreaks  
\begin{subequations}\label{eq:p10}
\begin{align}
    \min_{\substack{\mathbf{Q}, \boldsymbol{\zeta}, \boldsymbol{\alpha}, \\ \boldsymbol{\beta}, \boldsymbol{\gamma}, \boldsymbol{\eta}}} & \sum_{t=0}^T \Bigg(2L_{e} \Biggl[ \frac{u_0[t](2 - \Omega_0[t]) p_0[t]}{\zeta_0[t]}  \nonumber \\ 
    & \quad + \sum_{j=1}^K \frac{u_j[t](2 - \Omega_j[t]) p_j[t]}{\zeta_j[t]} \Biggr] + \omega \delta \widetilde{P}(\nu[t]) \Bigg) \tag{66}\\
    \text{s.t.} & \quad \zeta_k[t], \eta_k[t], \gamma_k[t] \geq 0, \quad \forall k \in \mathcal{K} \cup \{0\}, t,\\
    & \quad \frac{1}{g^2[t]} \le \chi^{lb}[t], \quad \forall t, \label{flight_bound} \\
    & \quad \eqref{p9-trans}-\eqref{p9-trans2},\; \eqref{p9-transa-1},\; \eqref{p9-transb-1},\;\eqref{flight1},\; \eqref{flight2}, \nonumber 
\end{align}
\end{subequations}
which can be solved by existing optimization tool box.
\subsection{Algorithm Design and Analysis}
The overall resource allocation framework for RSMA-enabled UAV probabilistic semantic communication is presented in Algorithm~\ref{alg:SCA_AO}. In the simulations, the stopping tolerance is set as $\varepsilon=10^{-4}$.
or when the maximum number of outer iterations $I_{\max}$ is reached.
\subsubsection{Convergence Analysis}
For Algorithm~\ref{alg:SCA_AO}, each subproblem is convex with respect to one optimization variable when the others are fixed, so the optimal solution of each subproblem can be obtained at every iteration. 
Moreover, the solutions obtained in one iteration remain feasible for the next iteration. 
As a result, the objective value of problem~(\ref{eq:p1}) is monotonically non-increasing over iterations. 
Furthermore, due to the transmit power and other system constraints, the objective function is bounded below. 
Therefore, according to the monotone bounded convergence (MBC) theorem, Algorithm~\ref{alg:SCA_AO} is guaranteed to converge within a finite number of iterations.

\begin{algorithm}[t]
\caption{Successive Convex Approximation-based Alternating Optimization Algorithm for Problem (\ref{eq:p1})}
\label{alg:SCA_AO}
\begin{algorithmic}[1]
\STATE \textbf{Input:} Initialized parameters $\mathbf{p}^{(0)}$, $\mathbf{W}^{(0)}$, $\boldsymbol{\Omega}^{(0)}$, $\mathbf{Q}^{(0)}$; maximum iteration number $I_2$; iteration index $i=0$.
\STATE \textbf{Output:} Minimal system energy consumption $E_{\min}$; optimal parameters $\mathbf{p}^*$, $\mathbf{W}^*$, $\boldsymbol{\Omega}^*$, $\mathbf{Q}^*$.

\WHILE{not converged \textbf{and} $i \le I_2$}
    \STATE Update $\mathbf{W}^{(i)}$ and $\mathbf{p}^{(i)}$ by solving problem (\ref{eq:p2}) through SCA with given $\boldsymbol{\Omega}^{(i-1)}$ and $\mathbf{Q}^{(i-1)}$.
    \STATE Update $\boldsymbol{\Omega}^{(i)}$ by solving problem (\ref{eq:p5}) using an optimization toolbox with given $\mathbf{Q}^{(i-1)}$, $\mathbf{W}^{(i)}$, and $\mathbf{p}^{(i)}$.
    \STATE Update $\mathbf{Q}^{(i)}$ by solving problem (\ref{eq:p7}) through SCA, where problem (\ref{eq:p10}) is constructed with given $\boldsymbol{\Omega}^{(i)}$, $\mathbf{W}^{(i)}$, and $\mathbf{p}^{(i)}$.
    \STATE Compute the objective value of problem (\ref{eq:p1}).
    \STATE Update $i = i + 1$.
\ENDWHILE%
\STATE \textbf{Return} $\mathbf{p}^*$, $\mathbf{W}^*$, $\boldsymbol{\Omega}^*$, $\mathbf{Q}^*$.
\end{algorithmic}
\end{algorithm}

\subsubsection{Complexity Analysis}
According to~\cite{RSMA5}, the complexity of obtaining the optimal solution to Problem~\eqref{eq:p2} is $\mathcal{O}(Y_1^2 Y_2 T)$, where $Y_1 = (L + 6)K + L + 2$ denotes the number of variables and $Y_2 = 12K + 5$ represents the number of constraints. Consequently, the total complexity for solving the joint power allocation and beamforming design subproblem is$:\mathcal{O}(I_1 L^2 K^3T)$, where $I_1$ denotes the number of iterations required by the SCA method.T
For the semantic compression ratio optimization subproblem, the interior-point method yields a complexity of $\mathcal{O}(K^3T)$.
For the UAV Trajectory optimization subproblem, the SCA yields a complexity of $\mathcal{O}(I_2 L^2 K^{2} T)$.
Therefore, the overall complexity of solving Problem~\eqref{eq:p1} is
$\mathcal{O}\left(I_1 I_2 L^2 K^3 T + I_2 K^3T+ I_2 L^2 K^{2} T\right) = \mathcal{O}(I_1 I_2 L^2 K^3 T)$, where $I_2$ represents the number of outer iterations in Algorithm \ref{alg:SCA_AO}.

The proposed SWEO algorithm can be extended to scenarios with mobile UEs when the UE trajectories are known or can be predicted over the mission duration. For a given UE mobility realization, the UE positions are treated as time-varying parameters rather than optimization variables; therefore, the monotone bounded convergence proof of SWEO remains valid for the resulting finite-horizon problem.

\section{Simulation Results and Analysis}
In this section, numerical results are provided to evaluate the effectiveness of the proposed SWEO algorithm for the RSMA-enabled UAV probabilistic semantic communication system. Unless otherwise specified, the main simulation parameters are listed in Table~\ref{tab4-1}.

\begin{table}[!t]
    \centering
    \caption{Simulation Parameters}
    \small 
    \begin{tabular}{|p{0.6\linewidth}|p{0.3\linewidth}|}
    \hline
    \textbf{Parameter Name}                & \textbf{Value}    \\
    \hline
    Flight speed                  & $V_{\max} = 20\, \text{m/s}$    \\
    \hline
    Radio bandwidth               & $B = 20\, \text{MHz}$ \\ 
    \hline
    Flight altitude               & $H = 100\, \text{m}$ \\
    \hline
    Number of UEs       & $3$ \\
    \hline
    Maximum transmit power   & $P=0.1\, \text{W}$ \\
    \hline
    Time slot length              & $\delta = 0.5\, \text{s}$ \\
    \hline
 Additional path loss for LoS&$\zeta^{\text{LoS}} = 1$~dB\\\hline
 Additional path loss for NLoS&$\zeta^{\text{NLoS}} = 20$~dB\\\hline
 Weight                       &$\omega$ = 0.01\\\hline
 Tip speed of the rotor blade &$U_{tip} = 120$~\text{m/s}\\\hline
 Rotor disc area & $A = 0.503$~$\text{m}^2$\\\hline
 Air density & $\rho = 1.225$~$\text{kg/m}^3$\\\hline
 Rotor solidity & $s=0.05$ \\\hline
 Fuselage drag ratio & $d_0 = 0.3$\\\hline
 Mean rotor induced velocity in hover & $v_0 = 4.03$ \\\hline
 Blade profile power in hovering status & $P_0 = 158.76\,~\text{W}$ \\\hline
 Induced power in hovering status & $P_i = 88.63\,~\text{W}$ \\\hline
 \multirow{2}{*}{Environment index} & $\eta = 9.6117$ \\ \cline{2-2}
                                   & $\gamma = 0.1581$ \\ \hline
\end{tabular}
    \label{tab4-1}
\end{table}

\begin{figure}[t] 
    \centering
    \includegraphics[width=0.9\columnwidth]{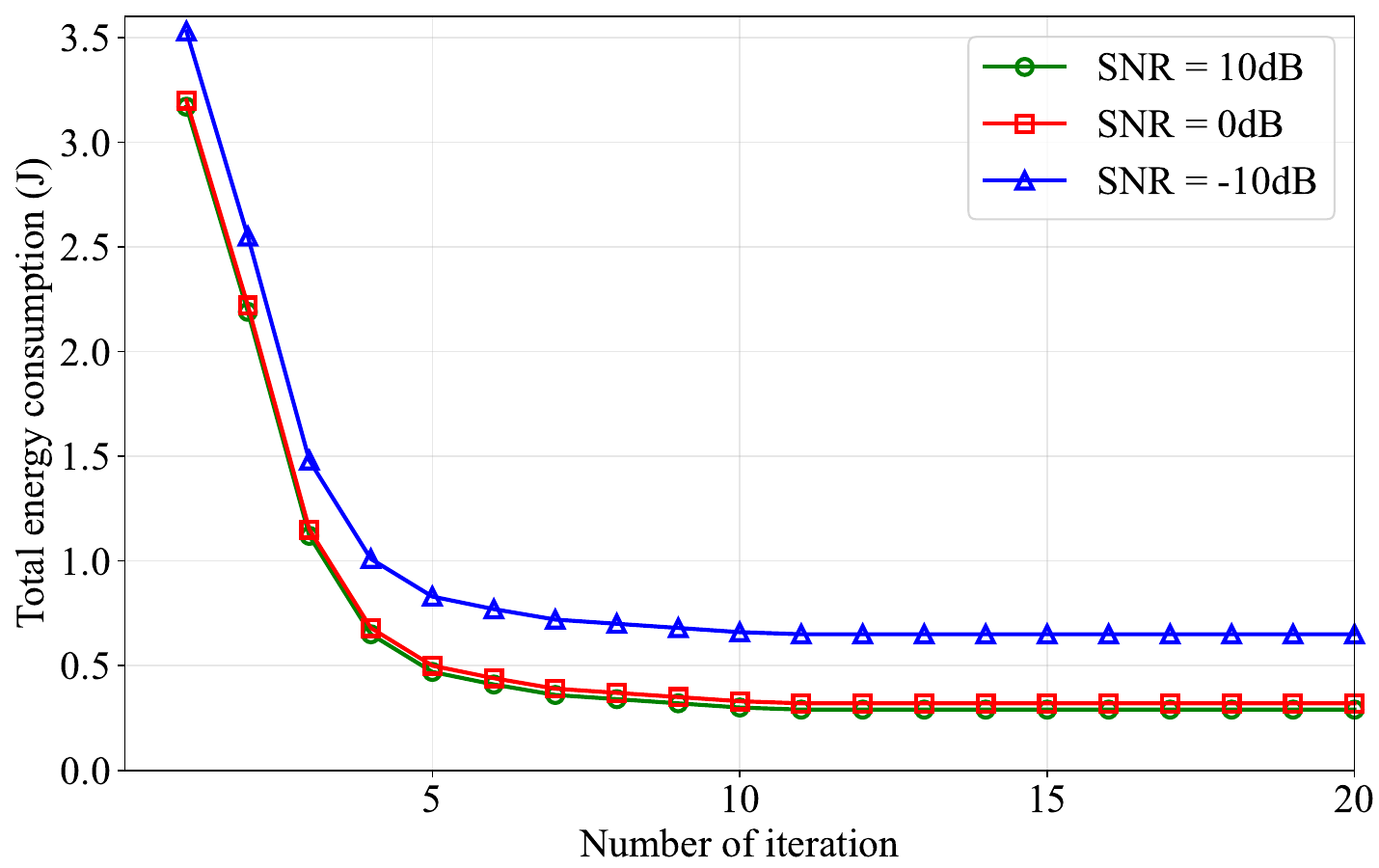} 
    \caption{The convergence of the proposed SWEO algorithm with different
SNR.}
    \label{fig:conver}
\end{figure}

Fig.~\ref{fig:conver} shows the convergence performance of the proposed SWEO algorithm under three different SNR. It can be observed that the total energy consumption decreases and eventually converges to a stable value within a finite number of iterations. Moreover, as the SNR decreases, the total energy consumption also converges to the higher stable values. This is because a lower SNR requires higher transmit power to satisfy the semantic accuracy and transmission constraints, which demonstrates the effectiveness of the proposed algorithm.

To evaluate the contribution of adaptive resource allocation, we compare the proposed SWEO algorithm with two benchmark schemes. The conventional scheme uniformly allocates the available resources among the common stream and private streams, while the random scheme randomly generates feasible resource-allocation variables. For the CR-related comparison, the conventional scheme corresponds to the case without PKG-assisted semantic compression.

\begin{figure}[t] 
    \centering
    \includegraphics[width=\columnwidth]{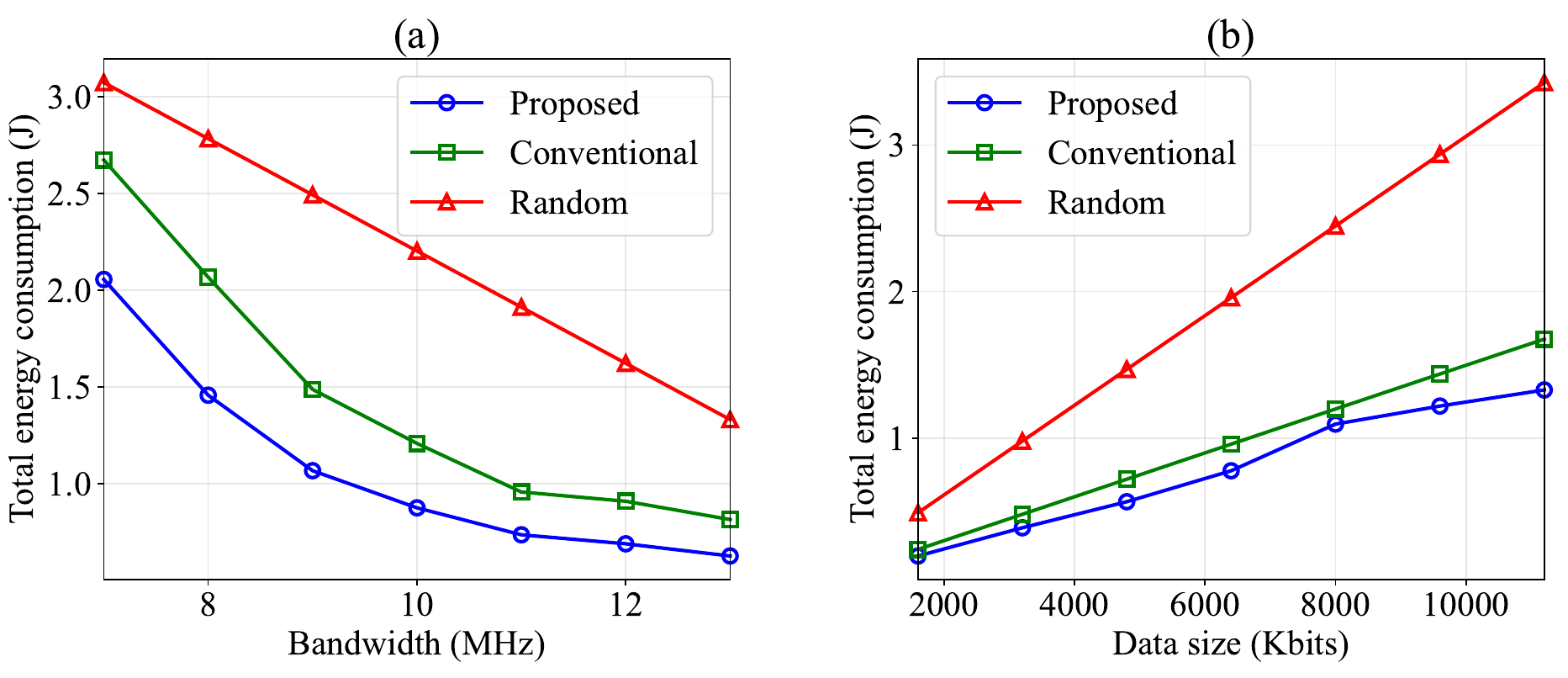} 
    \caption{Simulation results of power allocation and beamforming design subproblem.}
    \label{fig:5}
\end{figure}

Fig.~\ref{fig:5}(a) and Fig.~\ref{fig:5}(b) show the total energy consumption versus bandwidth and data size, respectively. As the bandwidth decreases or the data size increases, all schemes require higher energy consumption due to the increased semantic transmission time. The proposed SWEO algorithm consistently achieves the lowest energy consumption, because it adaptively coordinates RSMA resource allocation with the semantic payload and channel conditions. When the bandwidth decreases to $7$ MHz, the proposed scheme reduces the total energy consumption by $23.08\%$ and $33.09\%$ compared with the conventional and random schemes, respectively. When the data size increases, the corresponding energy savings reach up to $18.79\%$ and $60.27\%$, respectively.

\begin{figure}[t] 
    \centering
    \includegraphics[width=\columnwidth]{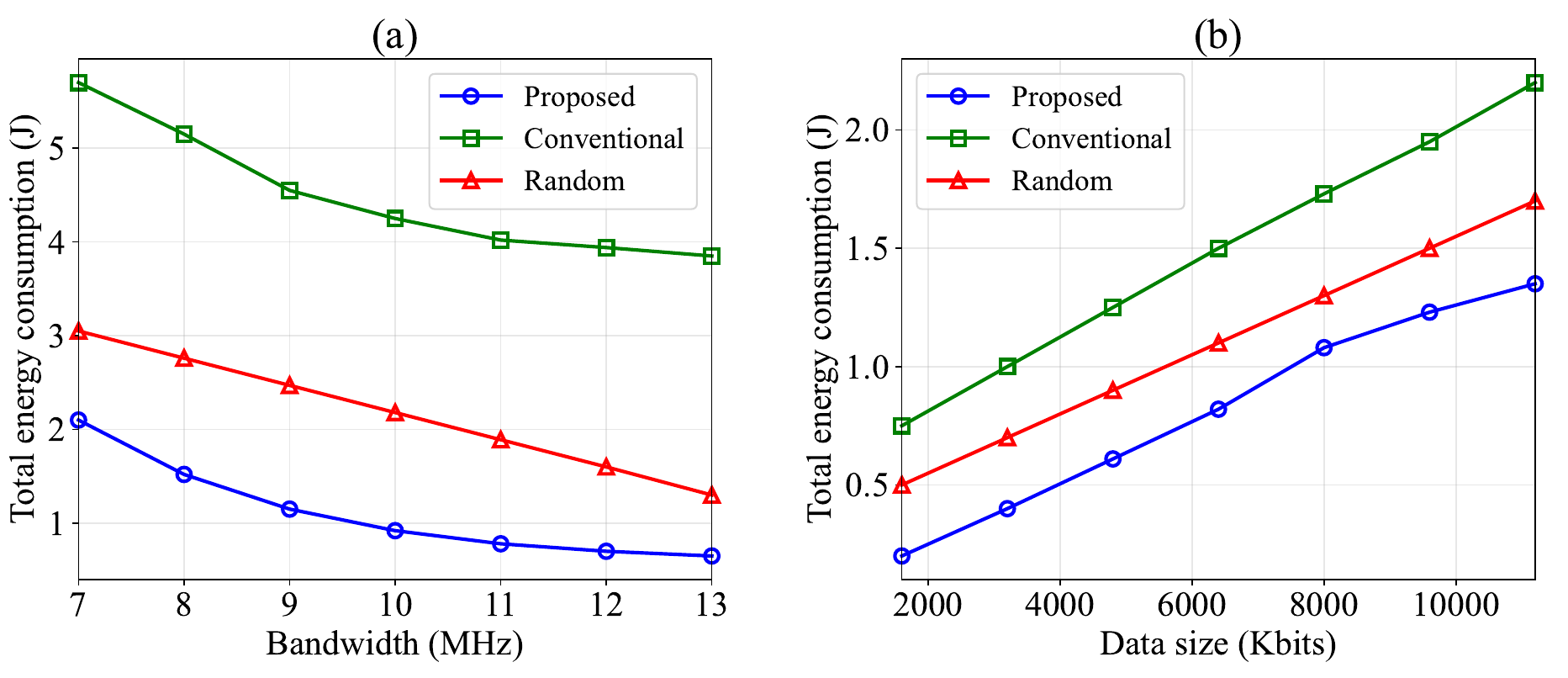} 
    \caption{Simulation results of compression ratio optimization subproblem.}
    \label{fig:6}
\end{figure}

\begin{figure}[t] 
    \centering
    \includegraphics[width=\columnwidth]{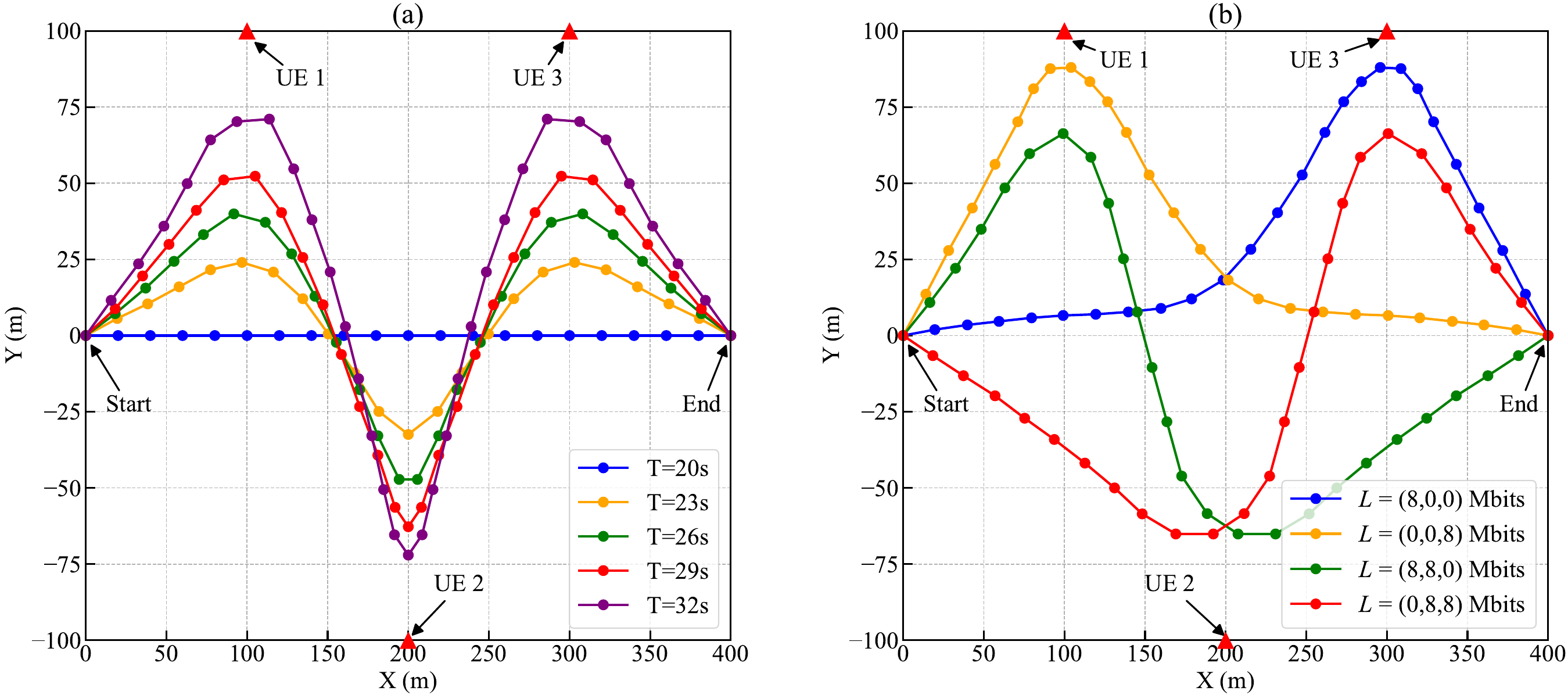} 
    \caption{Optimized UAV trajectories in different system settings: (a) Different flight time T under $L$ = (8, 8, 8) Mbits;
    (b) Different task requirement under $L$ under $T$ = 26s.}
    \label{fig:Trajectory}
\end{figure}

Fig.~\ref{fig:6} further evaluates the impact of CR optimization. Compared with the conventional scheme without semantic compression, the proposed method achieves lower energy consumption by adaptively balancing the computation overhead and the communication saving introduced by PKG-assisted compression. Specifically, increasing the CR reduces the semantic payload to be transmitted, but excessive compression may introduce additional computation cost and semantic accuracy loss. The proposed SWEO algorithm finds an energy-efficient CR under these coupled effects. It achieves up to $29.68\%$ and $63.84\%$ energy savings over the conventional and random schemes when the bandwidth varies, and up to $35.12\%$ and $53.30\%$ savings when the data size increases.

Fig.~\ref{fig:Trajectory}(a) illustrates the optimized UAV trajectories based on the proposed SWEO algorithm with different flight times for given task requirement $L = (8, 8, 8)$ Mbits. Since UEs have the same task requirement, the UAV approaches each TD in turn to provide communication service. This behavior is consistent with the intuition that minimizing the distance to UEs improves channel quality and ensures fair resource allocation.
Fig.~\ref{fig:Trajectory}(b) demonstrates the flight trajectory of the UAV under different task requirements. The trajectory adapts dynamically to the task distribution: the UAV flies closer to UEs with larger task requirement, while maintaining greater distance from those with minimal tasks. This demonstrates the algorithm’s ability to prioritize service based on task demand, thereby enhancing overall energy utilization. 

\begin{figure*}[t] 
    \centering
    \includegraphics[width=0.95\textwidth]{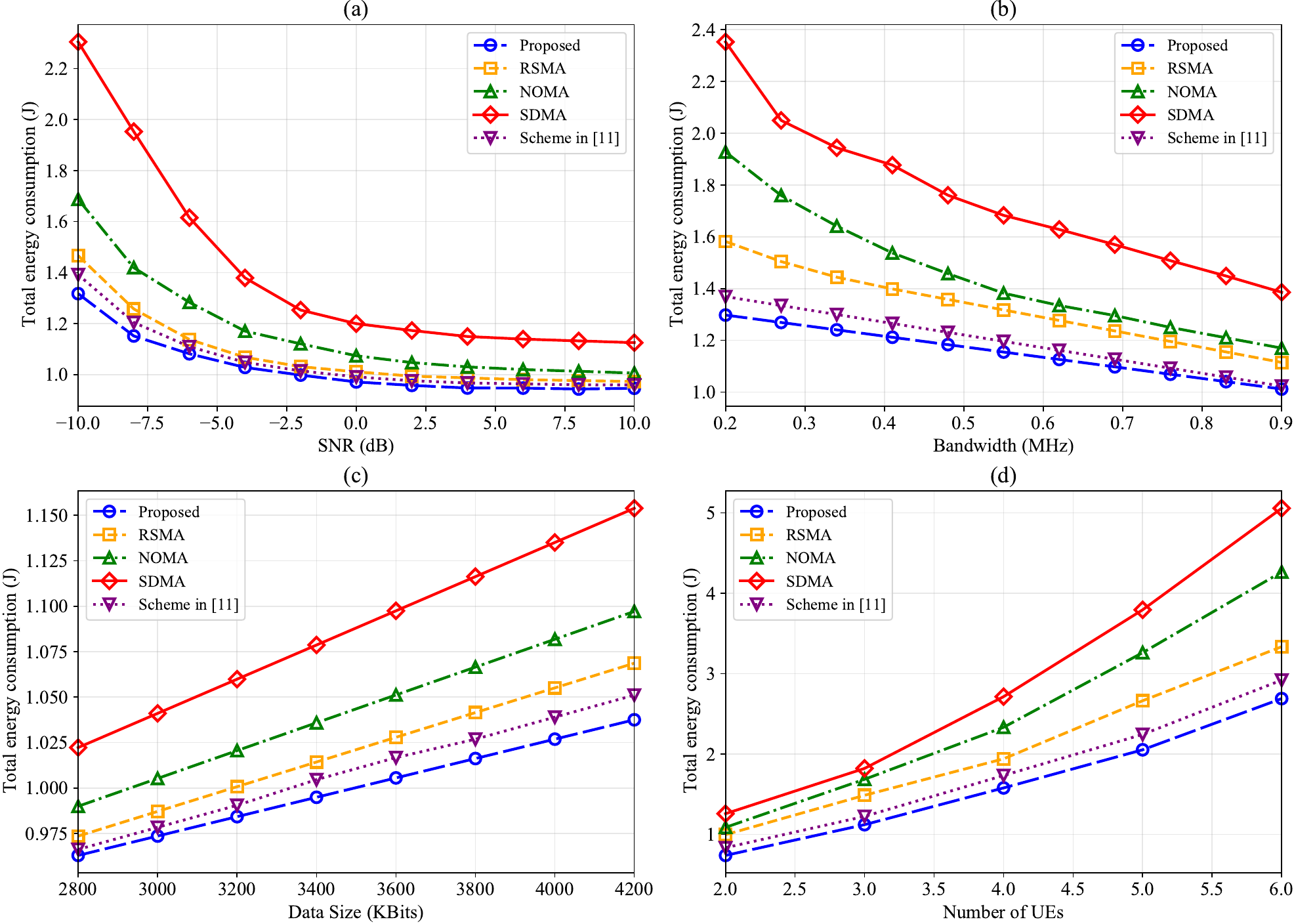} 
    \caption{Total energy consumption comparison under varying (a) SNR, (b) bandwidth, (c) data size, and (d) number of UEs.}
    \label{fig:Comparsion}
\end{figure*}

To further evaluate the effectiveness of the proposed UAV-enabled RSMA-PSC design, we compare the proposed SWEO algorithm with four benchmark schemes, including conventional RSMA, NOMA, SDMA, and the scheme in~\cite{RSMA5}. 
The RSMA, NOMA, and SDMA baselines do not employ the PKG-assisted PSC mechanism, while the scheme in~\cite{RSMA5} adopts the PSC-assisted RSMA transmission design but does not jointly optimize UAV trajectory in the proposed manner. 
For a fair comparison in the UAV-enabled scenario, all schemes are evaluated under the same task requirements, UAV mobility constraints, and channel settings.

Fig.~\ref{fig:Comparsion} presents the total energy consumption of different schemes under varying SNR, bandwidth, data size, and number of UEs. 
As shown in Fig.~\ref{fig:Comparsion}(a), the total energy consumption generally decreases as the SNR increases. 
This is because better channel conditions reduce the transmission effort required to satisfy the semantic transmission and delay constraints. 
The proposed scheme consistently achieves the lowest energy consumption across the entire SNR range. 
This demonstrates that the joint optimization of UAV trajectory, RSMA resource allocation, and semantic compression can better adapt to channel variations than the benchmark schemes.

Fig.~\ref{fig:Comparsion}(b) illustrates the total energy consumption with varying bandwidth. 
It shows that the total energy consumption decreases as the bandwidth increases, since a larger bandwidth shortens the transmission duration of semantic payloads and thus reduces the communication energy consumption. 
Compared with conventional RSMA, NOMA, and SDMA, the proposed UAV-enabled RSMA-PSC framework benefits from the semantic payload reduction introduced by PKG-assisted compression. 
Moreover, compared with the scheme in~\cite{RSMA5}, the proposed algorithm further reduces energy consumption by exploiting UAV trajectory adaptation, which improves the channel-dependent common and private transmission rates.

Fig.~\ref{fig:Comparsion}(c) illustrates the total energy consumption under different data sizes. 
As the data size increases, the total energy consumption of all schemes increases because more semantic information needs to be transmitted. 
However, the proposed scheme exhibits the slowest energy-growth trend and maintains the lowest energy consumption among all schemes. 
This is because the optimized CR reduces the communication load, RSMA efficiently delivers shared and private semantic components, and the UAV trajectory is adjusted to improve the air-to-ground channel conditions. 
In contrast, the conventional baselines suffer from larger communication payloads, while the scheme in~\cite{RSMA5} cannot fully exploit UAV mobility for energy-efficient semantic transmission.

Fig.~\ref{fig:Comparsion}(d) illustrates the total energy consumption under different numbers of UEs. As the number of UEs increases, the total energy consumption of all schemes increases due to the growing communication, computation, and transmission demands associated with serving more users. Nevertheless, the proposed scheme consistently achieves the lowest energy consumption and exhibits the slowest growth rate among all benchmark schemes. This demonstrates that the proposed UAV-enabled RSMA-PSC framework can effectively scale to larger multi-user scenarios.

\section{Conclusion}
This paper proposed a UAV probabilistic semantic communication framework for downlink transmission, leveraging RSMA to flexibly deliver shared and private semantic information.
A PKG-assisted semantic compression scheme was developed, where the semantic compression ratio serves as a tunable parameter to balance computation overhead and communication efficiency.
Accounting for the energy consumption of communication, computation, and UAV flight, we formulated a joint optimization problem over the UAV trajectory, power allocation, beamforming design, and semantic compression ratio.
To tackle the resulting non-convex problem, we proposed an SWEO algorithm based on alternating optimization, Lagrangian dual decomposition, and SCA. Furthermore, we proposed a novel weighted semantic accuracy metric that quantifies semantic distortion by prioritizing informative knowledge graph triples, which was incorporated as a quality-of-semantic-service constraint in the optimization framework.
Extensive numerical results verified the effectiveness of the proposed framework under dynamic UAV scenarios.
The proposed UAV-enabled RSMA-PSC framework consistently outperformed conventional multi-access schemes in terms of semantic performance across a wide range of system settings. 
Finally, it is worth noting that the assumptions of fixed computation capabilities and a single-UAV, fixed-altitude deployment simplify the onboard processing and propagation models, enabling us to focus on the coupled effects of UAV trajectory, power allocation, beamforming design, and compression ratio. Extending the proposed framework to jointly optimize dynamic computation frequency allocation, multi-UAV cooperation, and three-dimensional trajectory design with altitude optimization constitutes an interesting direction for future research in energy-constrained UAV networks.

\bibliography{reference}
\bibliographystyle{IEEEtran}
\bstctlcite{BSTcontrol}
\end{document}